\documentclass[a4paper,11pt]{article}

\usepackage{amsthm}
\usepackage{amssymb}
\usepackage{latexsym}
\usepackage{amsmath}
\usepackage{amsfonts}
\usepackage{version}
\usepackage[dvips]{graphicx}
\usepackage[utf8]{inputenc}

\usepackage[LGR,T1]{fontenc}%
\usepackage[francais,english]{babel}
\usepackage{url}
\usepackage{enumerate}
\usepackage{hyperref}
\usepackage{color}

\usepackage{fancyhdr}
\newcommand{\EE}{\mathbb{E}}
\newcommand{\RR}{\mathbb{R}}
\def\blue#1{\textcolor{black}{#1}}

\newtheorem{Theorem}{Theorem}[part]
\newtheorem{Definition}{Definition}[part]
\newtheorem{Proposition}{Proposition}[part]

\newtheorem{Lemma}{Lemma}[part]
\newtheorem{Corollary}{Corollary}[part]
\newtheorem{Remark}{Remark}[part]
\newtheorem{Example}{Example}[part]

\makeatletter \@addtoreset{equation}{section}

\@addtoreset{Definition}{section}

\@addtoreset{Theorem}{section}

\@addtoreset{Proposition}{section}

\@addtoreset{Property}{section}

\@addtoreset{Assumption}{section}

\@addtoreset{Corollary}{section}

\@addtoreset{Lemma}{section}

\@addtoreset{Remark}{section}

\@addtoreset{Example}{section}

\def\={\;=\;}

\def\be{\begin{eqnarray}}
\def\ee{\end{eqnarray}}
\def\b*{\begin{eqnarray*}}
\def\e*{\end{eqnarray*}}
\def\beq{\begin{equation}}
\def\eeq{\end{equation}}

\def\1{{\bf 1}}

\def \proof{{\noindent \bf Proof. }}
\def \eproof{\hbox{ }\hfill$\Box$}

\def \E{\mathbb{E}}

\def \R{\mathbb{R}}

\def\P{\mathbb{P}}

\def\cC{{\cal C}}

\def\cF{{\cal F}}

\def\cR{{\cal R}}

\def\ti{{t_i}}

\newcommand{\ud}
    {\mathrm{d}}

\newcommand{\esp}[1]
    {\ensuremath{%
     \mathbb{E}\!\!\left[#1\right]}}

\newcommand{\EFp}[2]
    {\ensuremath{
     \mathbb{E}_{#1}\!\!\left[#2\right] }}

\newcommand{\HP}[1] 
    {\ensuremath{\mathcal{H}^{#1}}}

\newcommand{\HPb}[1] 
    {\ensuremath{\mathcal{H}_\beta^{#1}}}

\newcommand{\Eb} 
    {\ensuremath{E_\beta}}

\newcommand{\ESP}[1] 
    {\ensuremath{\mathbb{S}^{#1}}}

\newcommand{\SPb}[1] 
    {\ensuremath{\mathcal{S}_\beta^{#1}}}

\newcommand{\Pot}[1] 
    {\ensuremath{\Pi_c^{#1}}}

\newcommand{\DM}[1] 
    {\ensuremath{\mathbb{D}^{1,#1}}}

\newcommand{\LM}[1] 
    {\ensuremath{\mathbb{L}_a^{1,#1}}}

\newcommand{\Lp}[1] 
    {\ensuremath{ \mathbf{L}^{#1} }}

\newcommand{\NH}[2] 
    { \ensuremath{ ||#2||_{\mathcal{H}^{#1}} } }

\newcommand{\NHb}[2] 
    { \ensuremath{ ||#2||_{\mathcal{H}_\beta^{#1}} } }

\newcommand{\set}[1]
    {\ensuremath{\{ #1 \}}}

\newcommand{\HYP}[1]
    {\ensuremath{({\bold H#1} ) }}

\newcommand{\Zr}{\ensuremath{ Z^{d\Re} }}

\newcommand{\ZriO}[1]{\ensuremath{ (Z^{d\Re})^{.i} }}

\newcommand{\ZrkiO}[1]{\ensuremath{ (Z^{d\Re})^{ki} }}

\newcommand{\tYr}{\ensuremath{\widetilde Y^{d\Re} }}

\newcommand{\tZr}{\ensuremath{\widetilde Z^{d\Re} }}

\newcommand{\hY}{\ensuremath{ \hat Y }}

\newcommand{\Xp} 
    {\ensuremath{X^{\pi} }}
\newcommand{\Xpt}[1] 
    {\ensuremath{X^{\pi}_{t_{#1}} }}

\newcommand{\Xpr}[1] 
    {\ensuremath{X^{\pi}_{r_{#1}} }}

\newcommand{\Ypb} 
    {\ensuremath{Y^{\pi,b} }}

\newcommand{\Ylb} 
    {\ensuremath{\tYr }}
\newcommand{\Ylbi} 
    {\ensuremath{\tilde Y^{b,i} }}
\newcommand{\Zlb} 
    {\ensuremath{\tZr }}

\newcommand{\dY} 
    {\ensuremath{\delta Y }}

\newcommand{\Ypt}[1] 
    {\ensuremath{Y^{\pi}_{t_{#1}} }}

\newcommand{\Ypr}[1] 
    {\ensuremath{Y^{\pi}_{r_{#1}} }}

\newcommand{\Ylp} 
    {\ensuremath{\widetilde{Y}^{\pi} }}

\newcommand{\Ylpt}[1] 
    {\ensuremath{\widetilde{Y}^{\pi}_{t_{#1}} }}

\newcommand{\Zlp} 
    {\ensuremath{\widetilde{Z}^{\pi} }}

\newcommand{\Zpt}[1] 
    {\ensuremath{Z^{\pi}_{t_{#1}} }}

\newcommand{\Zpr}[1] 
    {\ensuremath{Z^{\pi}_{r_{#1}} }}

\newcommand{\Zlpt}[1] 
    {\ensuremath{\widetilde{Z}^{\pi}_{t_{#1}} }}

\newcommand{\Zhpt}[1] 
    {\ensuremath{\widehat{Z}^{\pi}_{t_{#1}} }}

\newcommand{\Zbb} 
    {\ensuremath{\bar \Zr }}

\makeatletter
\renewenvironment{thebibliography}[1]{%
\list{\@biblabel{\@arabic\c@enumiv}}%
{\settowidth\labelwidth{\@biblabel{#1}}%
\leftmargin\labelwidth
\advance\leftmargin\labelsep
\@openbib@code
\usecounter{enumiv}%
\let\p@enumiv\@empty
\renewcommand\theenumiv{\@arabic\c@enumiv}}%
\sloppy
\clubpenalty4000
\@clubpenalty \clubpenalty
\widowpenalty4000%
\sfcode`\.\@m}
{\def\@noitemerr
{\@latex@warning{Empty `thebibliography' environment}}%
\endlist}
\makeatother

\renewenvironment{thebibliography}[1]{%
\begin{oldthebibliography}{#1}%
\setlength{\parskip}{0ex}%
\setlength{\itemsep}{0ex}%
}%
{%
\end{oldthebibliography}%
}

\title{An explicit Euler scheme with strong rate of convergence for financial SDEs with non-Lipschitz coefficients}

\author{
	Jean-Fran\c{c}ois Chassagneux
         \\\small Laboratoire  de Probabilit\'{e}s  et Mod\`{e}les  Al\'{e}atoires
         \\\small CNRS, UMR 7599,  Universit\'{e} Paris Diderot
         \\\small \sf jean-francois.chassagneux@univ-paris-diderot.fr
		\and
	       Antoine Jacquier
             \\\small  Department of Mathematics
             \\\small  Imperial College London
              \\\small  \sf a.jacquier@imperial.ac.uk
		\and
 		Ivo Mihaylov
             \\\small  Department of Mathematics
             \\\small  Imperial College London
              \\\small  \sf ivo.mihaylov06@imperial.ac.uk
              }

\begin{document}
\maketitle

\begin{abstract}
We consider the approximation of stochastic differential equations (SDEs) with non-Lipschitz drift or diffusion coefficients.
We present a modified explicit Euler-Maruyama discretisation scheme that allows us to prove strong convergence, with a rate. 
Under some regularity and integrability conditions, we obtain the 
optimal strong error rate. 
We apply this scheme to SDEs widely used in the mathematical finance literature, including the Cox-Ingersoll-Ross~(CIR), the $3/2$ and the Ait-Sahalia models,
as well as a family of mean-reverting processes with locally smooth coefficients.  
We numerically illustrate the strong convergence of the scheme and demonstrate its efficiency in a multilevel Monte Carlo setting. 
\end{abstract}

\vspace{5mm}


\vspace{5mm}

\noindent {\bf MSC Classification (2000):}  60H10, 65J15, 91G60.

 \newpage

\section{Introduction}

One of the main tasks in mathematical finance is to evaluate complex derivative products,
where the underlying assets are modelled by multi-dimensional SDEs, 
which rarely admit closed-form solutions.
Monte Carlo techniques are therefore needed to approximate these prices,
and Glasserman's book~\cite{glasserman2003monte} has become the main reference 
for a comprehensive overview of such methods with applications to financial engineering. 

Classical weak and strong convergence results for discretisation schemes of SDEs assume that the drift and the diffusion coefficients are globally Lipschitz continuous~\cite{kloeden1992numerical};
however many models used in the literature, such as the CIR, CEV, Ait-Sahalia models, violate this assumption. 
For pricing purposes, weak error is usually sufficient, 
but strong convergence rates are needed when using multilevel Monte Carlo methods~(MLMC), 
in order to optimise the computational complexity~\cite{giles2008multilevel,giles2009analysing}. 

In traditional Euler-Maruyama discretisation schemes, the approximation can potentially escape 
the domain of the true solution of the SDE.
In recent years, a lot of effort has focused on deriving schemes 
staying in restricted domains for SDEs with non-Lipschitz continuous coefficients~\cite{alfonsi2012strong, berkaoui2008euler, bossy2004, higham2002strong, hutzenthaler2012strong, neuenkirch2012first}. 
Several modifications have been introduced such as 
the drift-implicit~\cite{dereich2012euler} and the increment-tamed explicit Euler schemes~\cite[Theorem 3.15]{hutzenthaler2012numerical};
in the context of mathematical finance, a thorough overview of these can be found in~\cite{kloeden2012convergence}.

A now classical trick is to apply a suitable Lamperti transform in order to obtain
an SDE with constant diffusion coefficient, thereby translating all the non-smoothness to the drift.
In the context of non-globally Lipschitz coefficients, this idea, introduced by Alfonsi~\cite{alfonsi2005discretization}, was further exploited 
in~\cite{alfonsi2012strong,neuenkirch2012first}
to obtain strong $L^p$-convergence rates for implicit ``Lamperti-Euler'' schemes, 
in particular for the CIR and the Ait-Sahalia models, 
and for scalar SDEs with one-sided Lipschitz continuous drift and constant diffusion~\cite{neuenkirch2012first}.

Under sufficient differentiability conditions, modified It\^{o}-Taylor schemes~\cite{Jentzen2009} 
of order~$\psi>0$ provide pathwise convergence results of order $\psi-\varepsilon$ 
(for arbitrarily small~$\varepsilon>0$).
This approach relies on a localisation argument similar to that in~\cite{gyongy1998note}, 
with an auxiliary drift and a diffusion function chosen upon the discretised process exiting a sub-domain.
For irregular coefficients, some strong rates of convergence have been obtained
under more restrictive conditions in~\cite{gyongy1998note,gyongy2011note,yan2002,ngo2013strong}.

Motivated by these different approaches, our main contribution is to provide an efficient numerical
approximation of SDEs with non-globally Lipschitz coefficients.
We first present an explicit Euler scheme with a projection
for SDEs with locally Lipschitz and globally one-sided Lipschitz drift coefficient, which has a computational cost of the 
same order as the explicit Euler-Maruyama scheme. 
We prove strong rates of convergence for a wide family of SDEs, enlarging the range of parameters usually studied in explicit and implicit schemes. 
Under suitable assumptions, we are able to obtain fast convergence  
reaching the optimal rates of convergence.
The scheme shares some of the features of the tamed-scheme family. Its analysis however
does not require heavy technical tools. 
Having in mind applications to mathematical finance, 
the analysis is made for SDEs whose support is included in $(0,\infty)$. Nevertheless, the techniques
used here can be extended to the multi-dimensional case under some suitable assumptions.
An important contribution is to relate the choice of the scheme with the rate of explosion of the drift function at the boundaries of the domain through a locally Lipschitz continuous condition. 
To the best of our knowledge, thus far in the literature of \textcolor{black}{ tamed schemes, only 
the exploding behaviour at  one of the boundary  of the domain has been considered to obtain convergence rates. 
Our scheme considers generically both boundaries at the same time.}

We then turn our attention to SDEs with non-globally Lipschitz diffusion coefficients,
as often encountered in finance. We apply a Lamperti transformation to the process in order to shift 
the non-Lipschitz behaviour from the diffusion to the drift function, before using the modified scheme. 
This allows us to prove rates of convergence for the original process in the $L^{1+\varepsilon}$-norm for $\varepsilon\ge 0$; 
in particular, the rate of convergence for $\varepsilon=1$ can be used for MLMC applications, 
which we apply to the pricing of zero-coupon bonds and call spread options for correlated CIR processes.
\textcolor{black}{
In particular, we are able to prove convergence results for CEV/CIR-like model with non-constant smooth  coefficients, see Section \ref{sec FIX}.
Importantly, we also obtain new convergence results for the $3/2$-model, see Remark \ref{comp lit 3/2}. 
}

The remainder of the paper is structured as follows. 
In Section~\ref{sec:explicitscheme}, the modified Euler-Maruyama scheme is introduced, 
and the convergence results are proved in Section~\ref{sec:convergenceresult}. 
In Section~\ref{sec:applications}, the scheme is applied to families of SDEs, 
such as the CIR, the $3/2$ and the Ait-Sahalia models, widely used in mathematical finance, and the 
Ginzburg-Landau equation.
In Section~\ref{sec:numerics}, numerical results for the rates of convergence obtained are shown and discussed. 

\textbf{Notations}:
In the sequel, $D$  is the interval $(0,\infty)$.
We denote by $\bar{D}_\eta$ the domain $[\eta,\infty)$, and $\bar{D} := \bar{D}_0$. 
Furthermore, we define the interval $\breve{D}_\zeta:=(-\infty,\zeta]$ and $\check{D}_{\eta,\zeta}=\bar{D}_\eta\cap\breve{D}_\zeta$, for $\eta \le \zeta$.  
We denote by $\mathcal{C}^2(D)$ the space of twice differentiable functions with continuous derivatives on~$D$,
and by $\mathcal{C}_b^2(D)$ the space of functions in $\mathcal{C}^2(D)$ with first and second bounded derivatives.
We shall denote by $\mathbb{N}^+$ the set of strictly positive integers. 
\textcolor{black}{Given a probability space $(\Omega,\cF,\P)$, we denote $L^m(\Omega,\cF,\P)$, for $m>0$, the set of random variables~$Z$ such that $\|Z\|_m := \esp{|Z|^m}^{1/m} < +\infty$. In the sequel, we will simply write~$L^m$ when the probability space considered is clear from context.}
\blue{We denote by~$\EFp{t_i}{X}\equiv \esp{X | \mathcal{F}_{t_i}}$ the conditional expectation given the filtration $\mathcal{F}_{t_i}$ }

\section{Definitions and assumptions}\label{sec:explicitscheme}
Let $\left( \Omega, \mathcal{F}, (\mathcal{F}_t)_{t\geq0},\mathbb{P}\right)$ be a filtered probability space, 
and $W=(W_t)_{t\geq0}$ a one-dimensional standard $(\mathcal{F}_t)$-adapted Brownian motion. 
Consider a one-dimensional stochastic differential equation of the form
\begin{equation}\label{eqn:lampertitransformedsde}
\ud Y_t = f(Y_t) \ud t + \gamma(Y_t) \ud W_t , \qquad Y_0 = y_0 .
\end{equation}
Throughout this article, we shall assume the following:\\
{\HYP{y0}: the SDE~\eqref{eqn:lampertitransformedsde} admits a unique strong solution in $D=(0,\infty)$; 
the drift $f$ is locally Lipschitz continuous and globally one-sided Lipschitz continuous on $D$,
namely there exist $\alpha, \beta\geq 0$, $K>0$, such that for all $(x,y)\in D^2$:
}
\begin{align}
|f(x)-f(y)| &\le K\left(1 + |x|^\alpha + |y|^\alpha + \frac1{|x|^\beta} + \frac1{|y|^\beta}\right)|x-y|,
 \label{eq f loc lip}
 \\
 (x-y)\left(f(x)-f(y)\right) &\le K|x-y|^2; \label{eq f 1-sided lip}
\end{align}
furthermore, the diffusion function $\gamma$ is K-Lipschitz continuous on $\bar D$ for some $K>0$: for all $(x,y)\in \bar{D}^2$, the inequality $|\gamma(x)-\gamma(y)| \le K|x-y|$ holds.

\begin{Remark}
The function $\gamma$ could as well be defined on $D$. However, assuming the Lipschitz continuity of $\gamma$ on $D$ would lead to a natural extension of $\gamma$ on $\bar D$. 
\end{Remark}

\begin{Remark}
In many models used in practice (in particular the Feller/CIR diffusion in mathematical finance, see Section~\ref{sec:feller},
these assumptions are not met.
A suitable change of variables, however, allows us to bypass this: consider an SDE of the form
\begin{equation}\label{eqn:musigmasde}
\ud X_t = \mu(X_t) \ud t + \sigma(X_t) \ud W_t, 
\qquad  X_0 = x_0,
\end{equation}
where the process $X$ takes values in some domain $D_X\subseteq\mathbb{R}$.
\textcolor{black}{When $\sigma^{-1}$ is well defined and continuously differentiable on $D_X$,} 
the Lamperti transformation of $X$ is defined as 
$F(x) \equiv \int_{}^x \sigma(z)^{-1}\ud z$, and
It\^{o}'s Lemma implies that the process defined pathwise by $Y := F(X)$ satisfies~\eqref{eqn:lampertitransformedsde} with
$f\equiv F' \mu + \frac{1}{2} F'' \sigma^2$ and $\gamma \equiv F' \sigma$ is constant. 
\end{Remark}

Let $n\in\mathbb{N}^+$ be a fixed positive integer and $T>0$ a fixed time horizon.
Define the partition of the interval $[0,T]$ by $\pi := \{0=t_0<t_1<\ldots<t_n=T\}$,
with $\max_{i=0,\ldots,n-1} (t_{i+1}-t_i) =: h=\mathcal{O}(1/n)$.

For a closed interval $\cC\subset \RR$, we define $p_\cC:\R \to \cC$ as the projection operator onto~$\cC$.
For ease of notation, we let also $p_n:=p_{D_n}$, i.e. for $x \in \R$,
\begin{equation}\label{eqn:projection}
p_n(x) = \left\{
\begin{array}{lcll}
 n^{-k}\vee x \wedge n^{k'}&,& D_n=\check{D}_{n^{-k},n^{k'}}   &\text{ if } \alpha>0,\beta > 0
\\
 n^{-k}\vee x &,&  D_n= \bar{D}_{n^{-k}} &\text{ if } \alpha=0,\beta > 0
\\
  x \wedge n^{k'} &,& D_n= \breve{D}_{n^{k'}} &\text{ if } \alpha>0,\beta = 0
  \\
x  &,& D_n = \bar{D} &\text{ if } \alpha=\beta = 0  
\end{array}
\right. \,.
\end{equation}
In the following, we denote by $C$ a \textcolor{black}{positive} constant that depends only on $K$, $T$, $\alpha$, $\beta$, $y_0$, but whose value may change from line to line.
We denote it by $C_p$ if it depends on an extra parameter $p$.
We now introduce our explicit scheme for the discretisation process $\hat{Y}$:
\begin{Definition}\label{de scheme}
Set $\hY_0 := Y_0$ and for $i=0,\ldots, n-1$, 
$$
\hY_{t_{i+1}} := \hY_{t_{i}} + f_n(\hY_{t_i}) h_{i+1} + \bar{\gamma}(\hY_{t_i})\Delta W_{i+1},
$$
with $h_{i+1} := t_{i+1}-t_{i}$, $\Delta W_{i+1} := W_{t_{i+1}} - W_{t_i}$, $f_n := f \circ p_n$ and $\bar \gamma := \gamma \circ p_{\bar D}$.
\end{Definition}

\begin{Remark}\label{rem:dif schemes}
\label{re ext scheme}\text{}
\begin{enumerate}[(i)]
\item For some applications, it may be interesting to force the scheme to take values in a domain, e.g. intervals $\bar D$, $\bar{D}_\eta$, $\breve{D}_\zeta$ or even $\check{D}_{\eta,\zeta}$. 
To this end, we introduce some extensions of the previous scheme. 
For all $i \le n$, we define $\bar{Y}_{\ti{}} := p_{\bar D}(\hY_{\ti{}})$, $\tilde{Y}_{\ti{}} := p_{\bar{D}_\eta}(\hY_{\ti{}})$,  $\breve{Y}_{\ti{}} := p_{\breve{D}_\zeta}(\hY_{\ti{}})$
 and 
$\check{Y}_{\ti{}} := p_{\check{D}_{\eta,\zeta}}(\hY_{\ti{}})$, 
for some $\eta, \zeta>0$ to be determined later on, see Corollary~\ref{co conv ext scheme} for details.  
In Proposition~\ref{finite moments}, we prove finite moments and finite inverse moments for these modifications. 
\item Observe that for $\alpha=\beta=0$, $\hat{Y}$ is the usual Euler-Maruyama scheme, up to a projection onto $\bar{D}$.
\end{enumerate}
\end{Remark}

The following lemma shows how the properties of the initial drift~$f$ translate into the new projected drift~$f_n$ (proof in Appendix~\ref{app:proof lemma}): 
\begin{Lemma} \label{le fn lip}
For any $n\in\mathbb{N}^+$, 
the composition $f_n \equiv f\circ p_n$ is Lipschitz continuous 
with Lipschitz constant $L(n) = 2K (1+n^{k\beta}\1_{\set{\beta>0}}+n^{k'\alpha}\1_{\set{\alpha>0}})$, 
and one-sided Lipschitz continuous with the same Lipschitz constant~$K$ as that of~$f$.
\end{Lemma}

\begin{Remark} \label{re finite moments} 
For any $n\in\mathbb{N}^+$, since $f_n$ and $\gamma$ are Lipschitz continuous, 
an easy induction shows that the scheme in Definition~\ref{de scheme} satisfies 
$\max_{i=0,\ldots,n} \|\hY_{t_i}\|_2 < \infty$.
The bound is a priori non-uniform in~$n$, since the Lipschitz constant of $f_n$ depends on $n$.
\end{Remark}

We now introduce the following assumption, which implies that $L(n)^2 h \le C$, 
for all $n~\in~\mathbb{N}^+$,
and which relates the locally Lipschitz exponents~$\alpha$ and~$\beta$ 
to the size of the truncated domain~$D_n$:

\vspace{2mm}
\HYP{p}: the strictly positive constants $k$, $k'$ satisfy $2\beta k \le 1$ and $2\alpha k' \le 1$. 
\vspace{2mm}

We require additional assumptions to prove the strong convergence rate of our scheme: 
below \HYP{y1} imposes a condition on the moments of the process $Y$ in terms of the locally Lipschitz exponents $\alpha$ and $\beta$, to obtain a minimal convergence rate. 
We shall further impose regularity conditions on $f$ and $\gamma$ to obtain a better rate of convergence.

\vspace{2mm}

\HYP{y1}: \HYP{p} holds and there exist $q'>2(\alpha+1)$ and $q > 2\beta$ such that
\textcolor{black}{
$$ \sup_{t \in [0,T]}\esp{|Y_t|^{q'} + |Y_t|^{-q}} < \infty\;.$$ 
}
\vspace{2mm}

\HYP{y2}: \HYP{y1} holds, the drift function $f$ is of class $\mathcal{C}^2(D)$, 
and
\begin{align}\label{eqn hy2 ass}
\sup_{t \in [0,T]} \esp{ |\gamma(Y_t)f'(Y_t)|^2 + 
\left| f'(Y_t)f(Y_t) + \frac{\gamma^2(Y_t)}{2} f''(Y_t) 
\right|^2} <\infty.
\end{align}

For an implicit scheme, strong rates of convergence have been derived in~\cite{neuenkirch2012first}
assuming~\HYP{y2};
inspired by this paper, our motivation is to recover strong rates of convergence for the explicit scheme
in Definition~\ref{de scheme}.

\section{Convergence results}\label{sec:convergenceresult}
In this section we prove strong rates of convergence for the scheme in Definition~\ref{de scheme} under some of the assumptions stated above;
this result follows from estimates for the regularity of the processes $Y$ and $f(Y)$, and
the discretisation error of the scheme. 
\blue{Below, we give the results for the general case $\alpha,\beta \ge 0$, but in the
proof we restrict to the most complicated case $\alpha>0, \beta>0$.}
\subsection{Preliminary estimates}

Our first two results concern the error due to projecting the true solution $Y$ on $D_n$. 

\begin{Lemma} \label{le y-proj}
{
Assume that~\HYP{y0} and~\HYP{y1} hold. Then, for any $t\in[0,T]$, 
\begin{align*}
\esp{|Y_t-p_n(Y_t)|^2} \le C_{q,q'} \left( \frac1{n^{k(q+2)}}\1_{\set{\beta>0}}+ \frac1{n^{k'(q'-2)}}\1_{\set{\alpha>0}} \right)\,, 
\end{align*}
where $q,q'$ are given by \HYP{y1}.
}
\end{Lemma}

\proof
For any $t\in [0,T]$, we can write 
\begin{align*}
 \esp{|Y_t-p_n(Y_t)|^2} \le \frac1{n^{2k}}\P\left(Y_{t}<\frac1{n^k}\right)+\esp{|Y_t|^2\1_{\set{Y_t > n^{k'}}}}\;. 
\end{align*}
Set $\eta = q'/2$ and $\theta = q'/(q'-2)$, its conjugate exponent. 
H\"{o}lder's inequality yields
\begin{align*}
\esp{|Y_t|^2\1_{\set{Y_t > n^{k'}}}} \le \esp{|Y_t|^{q'}}^{1/\eta}\P\set{Y_t > n^{k'}}^{1/\theta}.
\end{align*}
Using \HYP{y1} and the set equality $\set{Y_t > n^{k'}}=\set{Y^{q'}_t > n^{k'q'}}$, Markov's inequality implies
$\esp{|Y_t|^2\1_{\set{Y_t > n^{k'}}}} \le C_{q'}n^{-k'(q'-2)}$. 
Likewise, since $\set{Y_{t}<n^{-k}}= \set{Y_t^{-q}>n^{kq}}$, Markov's inequality yields
$\P(Y_{t}<n^{-k}) \le C_q n^{-kq}$, and the lemma follows.
\eproof
\\

\begin{Lemma} \label{le f-fproj}
{
Assume that~\HYP{y0} and~\HYP{y1} hold.
Then, for any $t\in [0,T]$,  
\begin{align*}
\esp{|f(Y_t) -f_n(Y_t)|^2} \le 
C_{q,q'}\left(\frac{1}{n^{k(q-2(\beta-1))}}\1_{\set{\beta>0}}+ \frac{1}{n^{k'(q'-2(\alpha+1))}}\1_{\set{\alpha>0}}\right)=:K_2(n,q,q')\,,
\end{align*}
where $q,q'$ are given by \HYP{y1}.
}
\end{Lemma}

\proof
Using \eqref{eq f loc lip}, we observe that
\begin{align*}
|f(Y_t) -f_n(Y_t)|^2
 & \le C\left(1+ |Y_t|^{-2\beta}+|Y_t|^{2\alpha}\right)|Y_t-p_n(Y_t)|^2
\\
 & \le C\left(1+ |Y_t|^{-2\beta} \right)\frac1{n^{2k}}\1_{\set{Y_t < n^{-k}}}
 + C\left(1+ |Y_t|^{2\alpha} \right)|Y_t|^2 \1_{\set{Y_t > {n^{k'}}}} 
\\
& := A_1+A_2 .
\end{align*}
Set $\eta := q/(2\beta)$ and $\theta := q/(q-2 \beta)$. 
H\"{o}lder's inequality then yields
\textcolor{black}{
$$
\esp{A_1} \le\frac{C}{n^{2k}} \P\set{Y_t < 
n^{-k}} + \frac{C_q}{n^{2k}} \esp{|Y_t|^{-q}}^{1/\eta}\P\set{Y_t < 
n^{-k}}^{1/\theta},
$$
}
and \HYP{y1} {together with Markov's inequality} imply
$\esp{A_1} \le C_qn^{-k(q-2(\beta-1))}$.
Setting $\eta' := \frac{q'}{2(\alpha+1)}$ and 
$\theta' := \frac{q'}{q'-2(\alpha+1)}$, 
a similar computation gives
$\esp{A_2}~\le~C_{q'}n^{-k'(q'-2(\alpha+1))}$.
\eproof
\\

The following lemma provides a regularity result for the process~$Y$ 
and will be required for the main convergence result.  
For a given stochastic process~$X$ on $(\Omega, \mathcal{F}, (\mathcal{F}_t)_{t\geq 0}, \mathbb{P})$ 
and the partition $\pi$, 
we define its ``regularity'' by
\begin{equation}\label{eq:R}
 \cR_\pi[X] := \sum_{i=0}^{n-1} \int_{t_i}^{t_{i+1}}\esp{|X_t - X_{t_i}|^2} \ud t \;.
\end{equation}

\begin{Lemma} \label{le reg Y}
{
Assume that~\HYP{y0} and~\HYP{y1} hold.
The regularity of $Y$ satisfies $\cR_\pi[Y] \le C_{q,q'}h$, where $q,q'$ are given by \HYP{y1}.
}
\end{Lemma}

\proof
For $t\in (t_i,t_{i+1}]$, since $\gamma$ is $K$-Lipschitz,~\HYP{y1} implies
$$
\esp{|Y_t-Y_{t_i}|^2} \le C \esp{ \left(\int_{t_i}^t f(Y_s) \ud s\right)^2 + \int_{t_i}^t(|Y_s|^2+1) \ud s}
  \le Ch\left(1 + \frac1h\esp{ \left(\int_{t_i}^{t} f(Y_s) \ud s\right)^2}\right).
$$
For $t \in (t_i, t_{i+1}]$, by Lemma~\ref{le f-fproj}, we now compute
\begin{align*}
 \frac1h\esp{ \left(\int_{t_i}^{t} f(Y_s) \ud s\right)^2} &\le \esp{\int_{t_i}^{t_{i+1}} |f(Y_s)|^2 \ud s}
 \\
 &\le 2\left[ \int_{t_i}^{t_{i+1}} \esp{|f(Y_s)-f_n(Y_s)|^2} \ud s +\int_{t_i}^{t_{i+1}} \esp{|f_n(Y_s)|^2} \ud s \right]
 \\
 &\le Ch \left( K_2(n,q,q') + L(n)^2 \sup_{t \in [t_i,t_{i+1}]} \esp{1+|Y_t|^2}\;\right).
\end{align*}
Using \HYP{y1} and the inequality $L(n)^2h \le C$, which holds under $\HYP{p}$, we obtain
$\esp{|Y_t-Y_{t_i}|^2} \le C_{q,q'}h$ for $t\in(t_i,t_{i+1}]$, 
and the lemma follows from the upper bound
$$
\cR_\pi[Y] 
= \sum_{i=0}^{n-1} \int_{t_i}^{t_{i+1}}\esp{|Y_t - Y_{t_i}|^2} \ud t
 \le C \max\limits_{i=0,\ldots,n-1} \sup_{t\in[t_i,t_{i+1}]} \esp{|Y_t - Y_{t_i}|^2} \leq C_{q,q'}h\;.
$$
\eproof

We now compute upper bounds for the regularity of~$f(Y)$.
\begin{Lemma}\label{Lem regularity}
Assume that~\HYP{y0} and~\HYP{y1} hold.
\begin{enumerate}[(i)]
 \item Then
$\cR_\pi[f(Y)] \le C\left(K_2(n,q,q') +  L(n)^2h \right)$, where $q,q'$ are given by \HYP{y1}.
\item If moreover \HYP{y2} holds, then 
$\cR_\pi[f(Y)] \le Ch$.
\end{enumerate}
\end{Lemma}

\proof 
The inequality in~(i) is a direct consequence of the following computation:
\begin{align*}
\int_{t_i}^{t_{i+1}} \esp{|f(Y_t)-f(Y_{t_i})|^2} \ud t
 &  \le C\Big(  \int_{t_i}^{t_{i+1}} \esp{|f(Y_t)-f_n(Y_{t})|^2} \ud t \\
 & + \int_{t_i}^{t_{i+1}} \esp{|f_n(Y_{t})-f_n(Y_{t_i})|^2} \ud t \\
 &+ h \esp{|f_n(Y_{t_i})-f(Y_{t_i})|^2} \Big) \le C h\left(K_2(n,q,q') + L(n)^2 h\right),
\end{align*}
where we used Lemma~\ref{le f-fproj}. 
Let us now prove~(ii).
The drift function~$f$ is of class $\mathcal{C}^2(D)$ by~\HYP{y2},
and It\^{o}'s Formula on the interval $[t_i,t_{i+1}]$ implies
$$
f(Y_{t_{i+1}}) - f(Y_{t_i}) = \int_{t_i}^{t_{i+1}} \left(f'(Y_t)f(Y_t)+\frac12f''(Y_t)\gamma(Y_t)^2\right)\ud t
 + \int_{t_i}^{t_{i+1}} f'(Y_t)\gamma(Y_t) \ud W_t.
$$
Squaring and applying the Cauchy-Schwarz inequality the yields
$$
\esp{ |f(Y_{t_{i+1}}) - f(Y_{t_i})|^2}
 \le \int_{t_i}^{t_{i+1}}\esp{|\gamma(Y_t)f'(Y_t)|^2 + h\left|f'(Y_t)f(Y_t)+\frac{\gamma^2(Y_t)}{2}f''(Y_t)\right|^2} \ud t,
$$
and~(ii) follows from~\eqref{eqn hy2 ass}, direct integration on $[t_i, t_{i+1}]$ and summation.
\eproof

\subsection{Convergence result}
We consider here the discretisation error between the true process~$Y$ and the discretised process~$\hat{Y}$. 
Let us introduce the following notations:
\begin{align}\label{eq:delta}
\dY_{i} := Y_{t_i} - \hat{Y}_{t_i}, \qquad  \delta_n f_i 	:= f_n(Y_{t_i}) - f_n(\hat{Y}_{t_i}), \qquad \delta \gamma_i := \gamma (Y_{t_i}) - \bar \gamma(\hat{Y}_{t_i}) \;. 
\end{align}

The following key proposition provides a bound on the squared differences~$|\dY_{i}|^2$,
which depends on both the partition size and the regularity (in the sense of~\eqref{eq:R}), 
and which will be refined further below in Theorem~\ref{thm:Main}.

\begin{Proposition} \label{pr main error}
Assume that~\HYP{y0} and~\HYP{y1} hold, then
\begin{equation}\label{eq:GenUpBound}
\max_{i=0,\ldots,n} \esp{|\dY_{i}|^2}
 \le C\left(K_2(n,q,q') + \cR_\pi[f(Y)] + \cR_\pi[Y] \right)\,,
\end{equation}
where $q,q'$ are given by \HYP{y1}.
 \end{Proposition}
\proof
1.  We first show that the global error between the scheme and the solution is controlled by the sum of local truncation errors defined below.
Indeed, observe that
$$
Y_{t_{i+1}} = Y_{t_i} + f_n(Y_{t_i})h_{i+1}+ \bar \gamma(Y_{t_i})\Delta W_{i+1} + \zeta^d_{i+1} + \zeta^w_{i+1},
$$
for $i\leq n-1$, where
\begin{align*}
\zeta^d_{i+1} &:=  \int_{t_i}^{t_{i+1}} \left(f(Y_t)-f_n(Y_{t_i})\right) \ud t,\\
\zeta^w_{i+1} &:= \int_{t_i}^{t_{i+1}}\left(\gamma(Y_t)-\bar \gamma (Y_{t_i})\right)\ud W_t = \int_{t_i}^{t_{i+1}}\left(\gamma(Y_t)- \gamma(Y_{t_i})\right)\ud W_t.
\end{align*}
The last equality comes from the fact that $Y$ takes values in $D$ and $\bar \gamma (Y_{t_i})= \gamma (Y_{t_i})$, for all $i \le n$.
Therefore, squaring the difference $\dY_{i+1}$ gives
\begin{align}\label{eq dev square}
 |\dY_{i+1}|^2 =& |\dY_i|^2 + 2\delta Y_i \delta_n f_i h_{i+1} + 2\delta Y_i \delta \gamma_i \Delta W_{i+1}
 + 2 \delta Y_i \zeta^d_{i+1} + 2\delta Y_i \zeta^w_{i+1} 
 \\
 &+ | \delta_n f_i h_{i+1} +  \delta \gamma_i \Delta W_{i+1}
 +  \zeta^d_{i+1} +  \zeta^w_{i+1} |^2\;. \nonumber
\end{align}

Using the simple identity
$\EFp{t_i}{2\delta Y_i \delta \gamma_i \Delta W_{i+1} + 2\delta Y_i \zeta^w_{i+1}}=0$ and an application of Young's inequality
yields
\begin{align*}
 \esp{|\dY_{i+1}|^2} 
 & \le (1+C h) \esp{|\delta Y_i|^2}
+ C \esp{|\delta_n f_i h_{i+1}|^2 + |\delta \gamma_i|^2 h_{i+1} + \frac{|\EFp{t_i}{\zeta^d_{i+1}}|^2}{h} +  |\zeta^d_{i+1}|^2 
 + |\zeta^w_{i+1}|^2}\\
 & \le \left(1+C h + CL(n)^2h^2\right) \esp{|\dY_i|^2} + C \esp{ \frac{\left(\EFp{t_i}{\zeta^d_{i+1}}\right)^2}{h} + |\zeta^d_{i+1}|^2+ |\zeta^w_{i+1}|^2},
\end{align*}
since $f_n$ is one-sided Lipschitz continuous (Lemma~\ref{le fn lip}),
globally Lipschitz continuous with Lipschitz constant~$L(n)$ and $\gamma$ is Lipschitz continuous. 
Under~\HYP{p}, $L(n)^2h \le C$ and an iteration yields
\begin{align}
 \max_{i=0,\ldots,n} \esp{|\dY_{i}|^2} 
&\le  C \sum_{j=1}^n \esp{ \frac{\left(\EFp{t_j}{\zeta^d_{j}}\right)^2}{h} + |\zeta^d_{j}|^2 + |\zeta^w_{j}|^2} \label{eq conv stab fine}
\\
&\le  C \sum_{j=1}^n \esp{ \frac{|\zeta^d_{j}|^2}{h} + |\zeta^w_{j}|^2}. \label{eq conv stab}
\end{align}
2. We now provide explicit errors for the global truncation. 
As $\gamma$ is $K$-Lipschitz, we have 
$\esp{|\zeta^w_{i+1}|^2} \le C \int_{t_i}^{t_{i+1}}\esp{|Y_t-Y_{t_i}|^2} \ud t$,
and hence 
\begin{equation}\label{eq:RegBounds}
\sum_{i=1}^{n}\esp{|\zeta^w_{i}|^2} \le C  \cR_\pi[Y] .
\end{equation}
We now compute an upper bound for $\esp{|\zeta^d_{i+1}|^2}$.
Since 
\begin{equation}\label{eqn:zeta_d}
\zeta^d_{i+1}
 :=  \int_{t_i}^{t_{i+1}} (f(Y_t)-f_n(Y_{t_i})) \ud t = \int_{t_i}^{t_{i+1}} (f(Y_t)-f(Y_{t_i})) \ud t 
 + \int_{t_i}^{t_{i+1}} (f(Y_{t_i})-f_n(Y_{t_i})) \ud t,
\end{equation}
The Cauchy-Schwarz inequality yields
$$
\esp{|\zeta^d_{i+1}|^2}
 \le Ch\left(\int_{t_i}^{t_{i+1}} \esp{|f(Y_t)-f(Y_{t_i})|^2} \ud t + h \esp{|f(Y_{t_i})-f_n(Y_{t_i})|^2}\right),
$$
and Lemma~\ref{le f-fproj} implies
${\esp{|\zeta^d_{i+1}|^2}}
 \le Ch(\int_{t_i}^{t_{i+1}} \esp{|f(Y_t)-f(Y_{t_i})|^2} \ud t + h K_2(n,q,q'))$
and
$\frac1h\sum_{i=1}^n \esp{|\zeta^d_{i}|^2} \le C\left( K_2(n,q,q') + \cR_\pi[f(Y)]\right)$. 
Combining this with~\eqref{eq conv stab} and~\eqref{eq:RegBounds} concludes the proof.
\eproof
\vspace{2mm}

\textcolor{black}{
\begin{Remark}\label{re limit approach}
The method used to prove Proposition~\ref{pr main error} relies in particular on~\eqref{eq dev square}, 
which limits our result to the~$L^2$ setting.
We leave for further research the extension of this result to the $L^p$ setting, when $p > 2$. 
Obtaining such an extension would lead to a very
interesting improvement of the results in Section~\ref{sec:applications}.
\end{Remark}
}

We have kept the above result general, without \emph{a priori} assuming that the drift function belongs to $C^2(D)$. 
If we consider a constant diffusion and \HYP{y2}, we can  recover a better upper bound using~\eqref{eq conv stab fine}
instead of \eqref{eq conv stab} in the first part of the previous proof and prove a first-order strong rate of convergence. 
This will be illustrated in Proposition~\ref{first order rate} below. 

We now state the main result of our paper, namely a strong rate for $\delta Y_i$ defined in~\eqref{eq:delta}.
\begin{Theorem} \label{thm:Main}
{Assume that~\HYP{y0} holds}, then the inequality
\begin{align}
 \max\limits_{i=0,\ldots,n} \|\delta Y_{i}\|_2 \le C_{q,q'}h^r
\label{eq main res}
\end{align}
holds with $r=\min(\frac12-\frac{\beta}{q+2},\frac12-\frac{\alpha}{q'-2}) >0$ under 
\HYP{y1} 
by setting $(k,k') = (\frac1{q+2}, \frac1{q'-2})$ 
and $r= \min(\frac12, \frac{q+2}{4\beta}-\frac12,\frac{q'-2}{4\alpha}-\frac12) 
>0 $ under \HYP{y2} 
by setting $(k,k') = (\frac1{2\beta},\frac1{2\alpha})$.
\end{Theorem}

\proof
1. Assume \HYP{y1}. 
Combining Lemma~\ref{le reg Y} and Lemma~\ref{Lem regularity}(i) with \eqref{eq:GenUpBound} yields
\begin{align*}
 \max_{i=0,\ldots, n} \esp{|\dY_{i}|^2} & \le C(K_2(n,q,q') + L(n)^2h + h);
 \\
 &\le C_{q,q'}(h^{1-2 \beta k} + h^{k(q+2)-2\beta k} + h^{1-2 \alpha k' } + h^{k'(q'-2)-2\alpha k'} + h)\;.
\end{align*}
To balance the error terms, set $k=\frac1{q+2}$ and $k' = \frac1{q'-2}$, 
observing that under \HYP{y1}, \HYP{p} holds for this choice of parameters. 
Thus, we obtain $\max_{i=0,\ldots, n} \|\dY_{i}\|_2 \le  C_{q,q'} 
h^r$, with $r= \min(\frac12-\frac{\beta}{q+2},\frac12-\frac{\alpha}{q'-2})$, 
with $r>0$.

2.  Assume \HYP{y2}. 
Lemma~\ref{le reg Y} and Lemma~\ref{Lem regularity}(ii) with \eqref{eq:GenUpBound} imply
\begin{align*}
 \max_{i=0,\ldots, n} \esp{|\dY_{i}|^2} &\le C(K_2(n,q,q') + h)\;.
 \end{align*}
Setting $k=\frac1{2 \beta}$, $k' = \frac1{2 \alpha}$ yields $\max_{i=0,\ldots, 
n} \|\dY_{i}\|_2 \le  C_{q,q'} h^r$, 
where $r = \min(1/2, \frac{q+2}{4 \beta} - 1/2, \frac{q'-2}{4 \alpha} - 1/2)$.  
Since $\HYP{y2}$ implies $\HYP{y1}$, we observe that $r>0$.
\eproof
\vspace{2mm}

We now state the convergence results associated to the extensions of the scheme defined in Remark \ref{re ext scheme}.

\begin{Corollary} \label{co conv ext scheme}
{Assume that~\HYP{y0} holds.} 
Then the approximations~$(\tilde{Y}_{t_i})_{i\le n}$ and~$(\breve{Y}_{t_i})_{i\le n}$
defined in Remark~\ref{re ext scheme} satisfy
\begin{align*}
 \max\limits_{i=0,\ldots,n} \left( \|Y_{t_i}- \bar{Y}_{t_i}\|_2 + 
\|Y_{t_i}- \tilde{Y}_{t_i}\|_2+ 
\|Y_{t_i}- \breve{Y}_{t_i}\|_2 \right) \le 
C_{q,q'}h^r,
\end{align*}
{holds with $r=\min(\frac12-\frac{\beta}{q+2},\frac12-\frac{\alpha}{q'-2}) >0$ under 
\HYP{y1} 
by setting $(k,k') = (\frac1{q+2}, \frac1{q'-2})$ 
and $r= \min(\frac12, \frac{q+2}{4\beta}-\frac12,\frac{q'-2}{4\alpha}-\frac12) 
>0 $ under \HYP{y2} 
by setting $(k,k') = (\frac1{2\beta},\frac1{2\alpha})$, 
where $\eta := h^{2r/q}$ and $\zeta := h^{-2r/(q'-2)}$.}
\end{Corollary}

\proof
The proof follows by computing upper bounds for each of the three quantities on the left-hand side.
For all $i \le n$, since $p_{\bar D}$ is $1$-Lipschitz continuous, we can write
\begin{align*}
 \esp{|Y_{t_i}- \bar{Y}_{t_i}|^2} = \esp{|p_{\bar D}(Y_{t_i})- p_{\bar D}(\hY_{t_i})|^2} \le \esp{|Y_{t_i}-\hY_{t_i}|^2} = \EE\left|\delta Y_i\right|^2,
\end{align*}
and the upper bound for $\|Y_{t_i}- \bar{Y}_{t_i}\|_2$ follows from Theorem \ref{thm:Main}.

Set now $\eta = h^{2r/q}$. 
For $i \le n$, 
\begin{align}
 \esp{|Y_{t_i}- \tilde{Y}_{t_i}|^2} &\le 2 \left( \esp{|Y_{t_i}- p_{\bar{D}_\eta}(Y_{t_i})|^2} + \esp{| p_{\bar{D}_\eta}(Y_{t_i}) - p_{\bar{D}_\eta}(\hY_{t_i}) |^2} \right) \nonumber
 \\
 &\le  2 \left( \esp{|Y_{t_i}- p_{\bar{D}_\eta}(Y_{t_i})|^2} + \esp{| Y_{t_i} - \hY_{t_i} |^2} \right) \nonumber
\\
&\le C_{q,q'} \left(\esp{|Y_{t_i}- p_{\bar{D}_\eta}(Y_{t_i})|^2} + h^{2r} \right), \label{eq temp coro scheme}
\end{align}
where the last inequality follows from Theorem~\ref{thm:Main}.
A straightforward adaptation of the proof of Lemma~\ref{le y-proj} yields 
$\esp{|Y_{t_i}- p_{\bar{D}_\eta}(Y_{t_i})|^2} \le C_q \eta^{q}$, 
which gives the second bound.

Similarly, for $i \le n$, the equality $\E [|Y_{t_i} - p_{\breve{D}_\zeta}(Y_{t_i})|^2 ] = \E[ |Y_{t_i}-\zeta|^2 1_{\set{Y_{t_i}>\zeta}} ]$ holds, 
and an application of H\"{o}lder's inequality gives
$\E [|Y_{t_i} - p_{\breve{D}_\zeta}(Y_{t_i})|^2 ] \leq C_{q'} \zeta^{-(q'-2)}$. 
Choosing $\zeta=h^{-2r/(q'-2)}$ concludes the proof.
\eproof

\begin{Remark}\label{re tamed}
For SDEs defined on the whole real line, strong convergence rates have been proved 
using tamed explicit schemes~\cite{hutzenthaler2012strong,sabanis2013note}. 
The authors assumed that the drift satisfies~\eqref{eq f loc lip} and~\eqref{eq 
f 1-sided lip} with locally Lipschitz exponents $\alpha \in (0, \infty)$, 
$\beta=0$, $D=\R$
and that the diffusion is $K$-Lipschitz. 
Under these assumptions,~\eqref{eqn:lampertitransformedsde} has a unique strong 
solution~\cite{krylov1990simple}. 
Our modified scheme and a slight modification of the projection, namely,
$p_n(x) \equiv -n^{k'}\vee x \wedge n^{k'}$ can be applied to cover this case.
\end{Remark}

We now show that, as for the classical Euler scheme, 
our modified scheme may have a first-order strong rate of convergence if the diffusion coefficient is constant. 
This can be observed in practice, as shown in Section~\ref{sub se cir}. 
This also suggests that a similarly modified Milstein scheme, 
when the diffusion coefficient is not constant, will have a first-order strong rate of convergence.

\begin{Proposition}\label{first order rate}
{
Assume that $\gamma(x)\equiv\gamma>0$ for all {$x\in D$, and that \HYP{y0} with $q>6\beta-2$ and $q'>6\alpha+2$, and \HYP{y2} hold.} 
Then, $$\max_{i=0,\ldots,n} \left(\|\dY_{i}\|_2 + \|Y_{t_i}- \bar{Y}_{t_i}\|_2 + 
\|Y_{t_i}- \tilde{Y}_{t_i}\|_2+ 
\|Y_{t_i}- \breve{Y}_{t_i}\|_2 \right) \le C_{q,q'} h\,,$$
where we set  $\eta := h^{2/q}$ and $\zeta := h^{-2/(q'-2)}$ in the definition of $\tilde{Y}$ and $\breve{Y}$.
}
\end{Proposition}
\proof
The proof is similar to Step 2 in the proof of Proposition~\ref{pr main error}, 
but uses the sharper upper bound~\eqref{eq conv stab fine}. 
Since the diffusion function is constant, 
$\sum_{i=1}^{n}\esp{|\zeta^w_{i}|^2}$ is null, and using~\eqref{eqn:zeta_d} {and Lemma~\ref{le f-fproj}}, we can write
\begin{align}
& \max_i \esp{|\delta Y_i|^2}  \le \sum_{i=0}^{n-1}\esp{|\zeta^d_{i+1}|^2
+ \frac{(\EFp{\ti{}}{\zeta^d_{i+1}} )^2 } { h } } \label{eq jf temp 1}
\\
& \le  K_2(n,q,q') 
+ 
\sum_{i=0}^{n-1} \esp{\left|\int_{t_i}^{t_{i+1}}(f(Y_t)-f(Y_{t_i}))\ud 
t\right|^2
+\frac1h \left(\EFp{\ti{}}{\int_{t_i}^{t_{i+1}}(f(Y_t)-f(Y_{t_i}))\ud 
t}\right)^2
}. \nonumber
\end{align}
Moreover, It\^{o}'s Lemma implies
\begin{align*}
 \int_{t_i}^{t_{i+1}} (f(Y_t)-f(Y_{t_i})) \ud t &=  \int_{t_i}^{t_{i+1}} \left( 
\int_{{t_i}}^t f'(Y_u)f(Y_u)+\frac{1}{2}f''(Y_u)\gamma^2 \ud u+ \int_{{t_i}}^t 
f'(Y_u) \gamma \ud W_u\right) \ud t 
\end{align*}
which we can rewrite as
\begin{align*}
\int_{t_i}^{t_{i+1}} \left( \int_{{t_i}}^t 
f'(Y_u)f(Y_u)+\frac{1}{2}f''(Y_u)\gamma^2 \ud u \right) \ud t + 
\int_{t_i}^{t_{i+1}} (t_{i+1}-t) f'(Y_t) \gamma \ud W_t .
\end{align*}
Under \HYP{y2}, we then obtain easily, recalling~\eqref{eq jf temp 1}, that
\begin{align*}
\max_i \esp{|\delta Y_i|^2} \le C ( K_2(n,q,q') + h^2) \;.
\end{align*}
The proposition then follows by setting $(k,k')=(\frac1{2\beta},\frac1{2\alpha})$
and using the fact that $q>6\beta-2$ and $q'>6\alpha+2$, from Lemma~\ref{le f-fproj}.\\
The statement for $\|Y_{t_i}- \bar{Y}_{t_i}\|_2 $, $ \|Y_{t_i}- \tilde{Y}_{t_i}\|_2$, $\|Y_{t_i}- \breve{Y}_{t_i}\|_2$, 
follows from the same arguments as in Corollary \ref{co conv ext scheme}.
\eproof

\subsection{Moment properties of the schemes}
For later use, we show that our approximations have uniformly bounded second moments, 
which completes the result of Remark~\ref{re finite moments}.

\begin{Lemma}\label{lemma:boundmomteem1}
Assume that $\HYP{y0}$ and $\HYP{y1}$ hold.
Then, for $q,q'$  given by \HYP{y1},
$$\max_{i=0,\ldots,n} \esp{|\hY_{t_i}|^2 + |\bar{Y}_{t_i}|^2+ |\breve{Y}_{t_i}|^2+ |\tilde{Y}_{t_i}|^2} \le C_{q,q'}$$
with for $\breve{Y}$,   $\zeta := h^{-2r/(q'-2)}$ and for $\tilde{Y}$, $\eta := h^{2r/q}$,
recall Remark \ref{rem:dif schemes}, and with $r=\min(\frac12-\frac{\beta}{q+2},\frac12-\frac{\alpha}{q'-2}) >0$, 
under \HYP{y2}  $r= \min(\frac12, \frac{q+2}{4\beta}-\frac12,\frac{q'-2}{4\alpha}-\frac12) 
>0 $, and if moreover, $q > 6\beta -2$, $q'>6\alpha + 2$ and $\gamma(\cdot) \equiv \gamma>0$, $r=1$.
\end{Lemma}

\proof
Since $|\hat{Y}_i|^2 \leq 2 ( |Y_{t_i}-\hY_{t_i}|^2 + |Y_{t_i}|^2)$,
  \HYP{y1} and~Theorem~\ref{thm:Main} imply that
$$
\esp{|\hY_{t_i}|^2}
 \leq 2\left( \esp{|Y_{t_i}-\hY_{t_i}|^2} + \esp{|Y_{t_i}|^2} \right)
\leq C_{q,q'} (h^{2r}+1) \le C_{q,q'}
$$
holds for any $i\le n$, which proves the claim. \\
{
The statement for  $\bar{Y}$, $\breve{Y}$ and $\tilde{Y}$ follows from Corollary \ref{co conv ext scheme}
or Proposition \ref{first order rate}.
}
\eproof

\vspace{2mm}
We now consider the modifications $\tilde{Y}$ and $\breve{Y}$ defined in Remark~\ref{rem:dif schemes} 
and prove some finite moments or inverse moments for them, extending the previous result.

\begin{Proposition}\label{finite moments}
Assume that $\HYP{y0}$ hold and let $\zeta := h^{-2r/(q'-2)}$ and $\eta := h^{2r/q}$,
where $q$ and $q'$ are given by \HYP{y1}. 
\begin{enumerate}[(i)]
\item if \HYP{y1} holds, then 
$\max_{i=0,\ldots,n} \esp{\breve{Y}_{t_i}^{p}} \leq~C_{p,q,q'}$
for all $p\in [1,(q'-1) \vee 2]$;
\item if \HYP{y1} holds with $q\ge4$, then $\max_{i=0,\ldots,n} \esp{\tilde{Y}_{t_i}^{-p}} \leq~C_{p,q,q'}$
for all $p\in [1,q - 3 ]$.
\end{enumerate}
\end{Proposition}

{\proof
1. We first prove (i). We remark that the result for $p\in[1, 2]$ follows directly
from Lemma \ref{lemma:boundmomteem1}. We now assume that $1<p\le q-1$ and we introduce the sets \blue{$\breve{A}_i:=\set{Y_{\ti{}}\le\zeta}$
and $\breve{B}_i:=\set{|\delta\breve{Y}_{i}|> 1}$, where $\delta\breve{Y}_{i}:= \breve{Y}_{t_i} -Y_{t_i}$. We then observe that}
\begin{align*}
\blue{ \breve{Y}^p_{\ti{}}  = \breve{Y}^p_{\ti{}}\1_{\breve{A}_i^c}  +\breve{Y}^p_{\ti{}}\1_{\breve{A}_i \cap \breve{B}_i^c} +\breve{Y}^p_{\ti{}}\1_{\breve{A}_i \cap \breve{B}_i}}
\end{align*}
and deal which each terms in the right hand side separately.\\
Since \blue{$\breve{Y}_{t_i} \le \zeta$} by definition, we compute, for the first term,
\begin{align} \label{eq bound 1.a p moment}
\blue{ \esp{\breve{Y}^p_{\ti{}}\1_{\breve{A}_i^c}} \le \esp{{Y}^p_{\ti{}}} \le C_p\;.}
\end{align}
\blue{For the second term, as $|\delta\breve{Y}_{i}| \le 1$ on $\breve{B}_i^c$, we obtain}
\begin{align}\label{eq bound 1.b p moment}
\blue{ \esp{\breve{Y}^p_{\ti{}}\1_{\breve{A}_i \cap \breve{B}_i^c}} \le C_p(1+\esp{{Y}^p_{\ti{}}}) \le C_p.}
\end{align}
For the last term, we first  observe that for non negative $y$, $y'$ and $\theta\neq1$,
\begin{align}\label{eq fund theo calc}
(y')^\theta-y^\theta = \theta \int_0^1\left( (1-\lambda)y + \lambda y'\right)^{\theta-1} \ud \lambda (y'-y).
\end{align}
Using the above equality for $y' =  \breve{Y}_{\ti{}}$, $y'={Y}_{\ti{}}$ and $\theta=p$
we compute that
\begin{align*}
 |\breve{Y}^p_{\ti{}} - {Y}^p_{\ti{}}| \le C_p(\breve{Y}^{p-1}_{\ti{}} + {Y}^{p-1}_{\ti{}})|\blue{\delta\breve{Y}_{i}}|\;.
\end{align*}
Then since,
$
\blue{ \breve{Y}^p_{\ti{}}\1_{\breve{A}_i \cap \breve{B}_i} \le {Y}^p_{\ti{}} + |\breve{Y}^p_{\ti{}} - {Y}^p_{\ti{}}|\1_{\breve{A}_i \cap \breve{B}_i},}
$
we observe that 
\textcolor{black}{
\begin{align*}
 \esp{\breve{Y}^p_{\ti{}}\1_{\breve{A}_i \cap \breve{B}_i}} &\le C_p + C_p(1 + \zeta^{p-1})\esp{|\delta\breve{Y}_{i}|\1_{|\delta\breve{Y}_{i}|>1}}  \le C_p+C_p(1 + \zeta^{p-1})\esp{|\delta\breve{Y}_{i}|^2}
\end{align*}
}
Applying Corollary \ref{co conv ext scheme}, we thus obtain 
\begin{align}
 \esp{\breve{Y}^p_{\ti{}}\1_{A \cap B}} \le C_p(1 + \zeta^{p-1}h^{2r})\le C_p\;.
\end{align}
The proof of the first statement is concluded by combining the previous inequality
with \eqref{eq bound 1.a p moment} and \eqref{eq bound 1.b p moment}.
\\
2. We now prove (ii). We assume that $p \in [1,q-3]$ and that $q \ge 4$. \blue{We introduce
the set $\tilde{A}_i = \set{Y_{\ti{}} \ge \eta}$ and $\tilde{B}_i = \set{ |\delta\tilde{Y}_i |>\eta^2}$, where $\delta\tilde{Y}_i := \tilde{Y}_{t_i} - Y_{t_i}$.}
We observe that
\begin{align*}
\blue{ \tilde{Y}^{-p}_{\ti{}} = \tilde{Y}^{-p}_{\ti{}}\1_{\tilde{A}_i^c}  +\tilde{Y}^{-p}_{\ti{}}\1_{\tilde{A}_i \cap \tilde{B}_i^c} +\tilde{Y}^{-p}_{\ti{}}\1_{\tilde{A}_i \cap \tilde{B}_i}\;.}
\end{align*}
We are going to upper bound separately the expectation of each terms appearing in the right hand side of the above equality.\\
For the first term, since on $\tilde{A}_i^c$, $Y_{\ti{}}\le \tilde{Y}_{\ti{}}$ holds by definition, we get
\begin{align*}
\blue{ \esp{\tilde{Y}^{-p}_{\ti{}}\1_{\tilde{A}_i^c} } \le  \esp{{Y}^{-p}_{\ti{}}\1_{\tilde{A}_i^c} } \le C_p\;.}
\end{align*}
For the second term, observing that
$
  \frac1{Y_{\ti{}}} -  \frac1{\tilde{Y}_{\ti{}}} = \frac{\delta\tilde{Y}_i }{Y_{\ti{}} \tilde{Y}_{\ti{}}}
$, by \blue{\HYP{y1} we compute }
\begin{align*}
\blue{ \esp{\tilde{Y}^{-p}_{\ti{}}\1_{\tilde{A}_i \cap \tilde{B}_i^c}} }& \le C_p\esp{{Y}^{-p}_{\ti{}} + \left|\frac{\delta\tilde{Y}_i }{Y_{\ti{}} \tilde{Y}_{\ti{}}}\right|^{p}\1_{\tilde{A}_i \cap \tilde{B}_i^c}}
  \le C_p,
\end{align*}
\blue{since on $\tilde{A}_i\cap \tilde{B}_i^c$, $|\delta\tilde{Y}_i | \le \eta^2$ and $\frac1{{Y}_{\ti{}}} \le \frac1\eta$.}
For the last term, we compute that
\begin{align*}
\blue{\esp{\tilde{Y}^{-p}_{\ti{}}\1_{\tilde{A}_i \cap \tilde{B}_i}} \le C_p\esp{ {Y}^{-p}_{\ti{}} + |\tilde{Y}^{-p}_{\ti{}}-{Y}^{-p}_{\ti{}}|\1_{\tilde{A}_i \cap \tilde{B}_i}}}
\end{align*}
and using \eqref{eq fund theo calc}, we get
\begin{align*}
\esp{\tilde{Y}^{-p}_{\ti{}}\1_{\tilde{A}_i \cap \tilde{B}_i}} & \le C_p(1 + \esp{ (\tilde{Y}^{-p-1}_{\ti{}}+{Y}^{-p-1}_{\ti{}})|\delta\tilde{Y}_i |\1_{\tilde{A}_i \cap \tilde{B}_i} } \\
& \le  C_p(1+\eta^{-(p+1)})\esp{|\dY_{\ti{}}|\1_{\set{|\delta\tilde{Y}_i |> \eta^2}}}\;.
\end{align*}
Using the Cauchy-Schwarz inequality and then applying Chebyshev's inequality, we obtain
\begin{align*}
\blue{\esp{\tilde{Y}^{-p}_{\ti{}}\1_{\tilde{A}_i \cap \tilde{B}_i}}  \le C_p(1+ \eta^{-(p+3)})h^{2r} \le C_p\;,}
\end{align*}
which concludes the proof for this step.
\eproof
}

\section{Applications}\label{sec:applications}
\textcolor{black}{As a first illustration,} we now apply our results to various stochastic differential equations widely used in the literature.
\subsection{CIR model}\label{sec:feller}
We consider the Feller diffusion~\cite{feller1954}, defined as the unique strong solution to
\begin{equation}\label{eq:FellerSDE}
\ud X_t = \kappa(\theta-X_t) \ud t + \xi \sqrt{X_t} \ud W_t, \qquad X_0 = x_0 > 0 ,
\end{equation}
where $W$ is a Brownian motion, and $\kappa$, $\theta$, $\xi$ are strictly positive constant parameters.
This process is widely used in mathematical finance, 
both for interest rate modelling~\cite{cir1985} and for the instantaneous variance of a stock price process~\cite{heston1993}. 
Under the Feller condition $\omega:=2\kappa\theta/\xi^2>1$, $X$ remains strictly positive almost surely,
and It\^o's Lemma implies that the Lamperti transform $Y = \sqrt{X}$ satisfies
\begin{equation}\label{eqn:lamperticir}
\ud Y_t= f(Y_t) \ud t + c\, \ud W_t, \qquad Y_0 = \sqrt{x_0} > 0, 
\end{equation}
where 
\begin{equation}\label{eqn:lamperticir2}
 f(x) \equiv a/x + b x,
\qquad
a := (4\kappa\theta - \xi^2)/8, 
\qquad
b := - \kappa/2 , 
\qquad
c := \xi/2;
\end{equation}
furthermore, $a > 0$ when the Feller condition holds.
Since $X = Y^2$, proving a rate of convergence for a discretisation scheme for the process $Y$ 
will allow us to obtain a rate of convergence for the process $X$.
In the following corollary, we apply Theorem~\ref{thm:Main}
to provide bounds for $\|\dY_{i}\|_2$ and $\|\delta X_i\|_1$, where 
$\delta X_i := X_{t_i} - \hat{X}_{t_i} = Y_{t_i}^2 - \hat{Y}_{t_i}^2$.

\begin{Corollary}\label{cor:circonvergencenew}
For $\omega >2$,
$\max_{i=0,\ldots,n} \left(\|\delta Y_{i} \|_2 + \|\delta X_{i} \|_1\right) \le { C_{r} }h^r$
holds, where
\begin{equation}\label{eqn:cir_r}
\left\{\begin{array}{ll}
\displaystyle r\in \left({\frac16},\frac{1}{2}-\frac{1}{\omega+1}\right),\qquad & \text{if }2<\omega\leq 3,\\
r=1/2,\qquad & \text{if }3< \omega \leq 5,\\
r=1,\qquad & \text{if }\omega>5.
\end{array}
\right.
\end{equation}
\end{Corollary}

\proof
Consider first the bound for $\|\delta Y_{i} \|_2$.
The drift of $Y$ is one-sided Lipschitz continuous and locally Lipschitz continuous with exponents $\alpha=0$ and $\beta=2$, and the diffusion is constant, hence Lipschitz continuous. 
From~\cite[~page~5]{dereich2012euler}, we know that 
$\sup_{t\in [0,T]} \mathbb{E} ( |X_t|^p ) < +\infty$  for all $p > -2 \kappa \theta/\xi^2$, 
and therefore 
\begin{align}\label{eq bound moment y}
\sup_{t\in [0,T]} \mathbb{E} (|Y_t|^{-\ell}) < +\infty \;\text{ for all }\;
\ell<4\kappa\theta/\xi^2=2\omega.
\end{align}
In the case $2<\omega\leq3$, we choose $q \in (4,2\omega)$ and fix~$k=1/(q+2)$, so that \HYP{p} holds 
(no condition on $k'$ is required since $\alpha=0$)
and \HYP{y1} holds as well.
From Theorem~\ref{thm:Main} it follows that the convergence rate is given by $r:=1/2-\beta/(q+2)$. We compute easily, since $\beta=2$, that
$r \in (\frac16,\frac{1}{2}-\frac{1}{\omega+1})$, depending on the choice of $q \in (4,2\omega)$.\\
Consider now the case $3<~\omega$.
We compute that 
$\mathbb{E}( |f(Y_t) f'(Y_t) + \frac{1}{2} c^2 f''(Y_t)|^2) \leq C\mathbb{E} ( |Y_t|^2 + |Y_t|^{-6} ) \leq C$ hold. 
{
Combining the previous inequality with \eqref{eq bound moment y}, we obtain that~\HYP{y2} holds.
{If $3<\omega\leq5$,} fix $q \in (6,2\omega)$ and set $k=1/4$, it follows  that $r=\min(1/2,(q+2)/8-1/2)=1/2$ from Theorem~\ref{thm:Main}.
}
The case $\omega >5$ follows directly from  Proposition~\ref{first order rate}, {since there exists a $q\in(10,2\omega)$ such that $\esp{Y_t^{-q}} < \infty$ for all $t\in[0,T]$ by~\eqref{eq bound moment y}}.

We now prove the corollary for the difference $\delta X_i$.
The Cauchy-Schwarz inequality and the result above imply
\begin{align*}
\mathbb{E} [|\delta X_i | ]
 &= \mathbb{E} \left[|(Y_{t_i} -  \hat{Y}_{t_i}) (Y_{t_i} +  \hat{Y}_{t_i})| 
\right]
\leq \sqrt{\mathbb{E} (|\dY_i|^2) \mathbb{E} \left[|Y_{t_i} + 
\hat{Y}_{t_i}|^2\right] } \\
&\leq C_rh^{r} \sqrt{\mathbb{E} ( |Y_{t_i}|^2 ) + \mathbb{E} ( |\hat{Y}_{t_i}|^2 
)}
\leq C_rh^{r} ,
\end{align*}
since $\mathbb{E}(|Y_{t_i}|^2)$ and $\mathbb{E}(|\hat{Y}_{t_i}|^2)$ 
are finite from~\cite[Lemma 
3.2]{higham2002strong} and Lemma~\ref{lemma:boundmomteem1}.
\eproof
\vspace{2mm}

{Define $\delta \breve{X}_i := X_{t_i} - \breve{X}_{t_i}$, where $\breve{X}_{t_i} :=  \breve{Y}_{\ti{}}^2$,
recall Remark~\ref{rem:dif schemes}.} 
We now consider a general $L^{1+\varepsilon}$-norm for convergence of the discretisation scheme of process~$X$.
{
\begin{Corollary}\label{cor:cir L1eps norm}
Suppose that $\omega>2$ and fix $\varepsilon \ge 0$. 
Then
$$\max_{i=0,\ldots,n} \|\delta \breve{X}_{i} \|_{1+\varepsilon} \le { C_{r,\varepsilon} }h^{r/(1+\varepsilon)},$$
with $r$ defined as in~\eqref{eqn:cir_r} and where we set $\zeta := h^{-\frac{2r}{q'-2}}$, with $q'=3+4\epsilon$ in the definition of $\breve{X}=\breve{Y}^2$ in Remark~\ref{rem:dif schemes}.
\end{Corollary}
}
\proof
For all $i\geq0$, we have
\begin{equation*}
\begin{array}{rl}
\|\delta \breve{X}_{i} \|_{1+\varepsilon}^{1+\varepsilon} 
  & = \esp{|X_{t_i} - \breve{X}_{t_i}|^{1+\varepsilon}}
  = \esp{|Y_{t_i} - \breve{Y}_{t_i}| |Y_{t_i} - \breve{Y}_{t_i}|^\varepsilon |Y_{t_i} + \breve{Y}_{t_i}|^{1+\varepsilon} } \\
  & \leq \|Y_{t_i} - \breve{Y}_{t_i} \|_2   \sqrt{\esp{\left(|Y_{t_i}| + |\breve{Y}_{t_i}|\right)^{2+4\varepsilon}}}\,.
\end{array}
\end{equation*}
From \eqref{eq bound moment y}, we have that $\esp{|Y_{t_i}|^{2+4\varepsilon}} < C_\epsilon$. 
Similarly, since $\esp{|Y_{t_i}|^{q'}} < +\infty$, we obtain from Proposition~\ref{finite moments}(i), that
$\esp{|\breve{Y}_{t_i}|^{2+4\varepsilon}} < C_{r,\varepsilon}$. 
This moment bounds, combined with Corollary~\ref{co conv ext scheme} (or Proposition \ref{first order rate}, when $r=1$)  and the above inequality, leads to $\|\delta \breve{X}_{i} \|_{1+\varepsilon}^{1+\varepsilon}  \le C_{r,\varepsilon} h^r$.
\eproof

\begin{Remark}
\textcolor{black}{
To the best of our knowledge, the best convergence result in term of range for the parameter $\omega$ are obtained using an implicit Euler scheme,
in \cite{hutzenthaler2014strong}, see also the references therein. In this paper, $\omega$ belongs to $(0.5,\infty)$ whereas our results are valid for $\omega \in (2,\infty)$. The main advantage of our scheme is its explicit nature that
allows to retrieve convergence results for non-constant coefficients as illustrated in the next section. Let us also mention in this regard the very recent paper \cite{bossy2015strong} on the symmetrised Milstein scheme.
%
}
\end{Remark}

\subsection{Locally smooth coefficients}\label{sec FIX}
We now consider a stochastic differential equation of the form~\eqref{eqn:musigmasde},
with drift function $\mu(x) \equiv \mu_1(x)-\mu_2(x)x$, where 
$\mu_1, \mu_2 : D \to \R$, and diffusion function $\sigma(x) \equiv \gamma x^\nu$, with $\gamma>0$ and
$\nu\in [1/2,1]$.
This model encompasses the Feller diffusion (see Section~\ref{sec:feller}) and the CEV model~\cite{coxross1976},
both widely used in mathematical finance. 
For the special case $\nu=1$, the diffusion function is $K$-Lipschitz and our scheme applies directly to the process $X$ as long as~\eqref{eq f loc lip} and~\eqref{eq f 1-sided lip} hold for the drift function $\mu$.

We now focus on the case~$\nu\in[1/2,1)$. 
The Lamperti transform reads $F(x) \equiv \int^x \ud y /\sigma(y) \equiv  \frac{1}{\gamma (1-\nu)} x^{1-\nu}$,
with inverse $F^{-1} (y) \equiv \left[\gamma(1-\nu) y \right]^{\frac{1}{1-\nu}}$.
The process~$Y=F(X)$ is the solution to $\ud Y_t = f(Y_t) \ud t + \ud W_t$, with $Y_0 = F(x_0)$ and
\begin{equation}\label{eqn:flocallysmooth}
f(y) \equiv \frac{\mu\left(F^{-1}(y)\right)}{\sigma\left(F^{-1}(y)\right)}-\frac{1}{2} \sigma'\left( F^{-1}(y) \right).
\end{equation} 
In order for the functions $\mu$ and $\sigma$ to satisfy the required conditions, we assume:\\
\HYP{s0}: $\nu \in[1/2,1)$, and $\mu_1, \mu_2$ are bounded, belong to $\mathcal{C}_b^2(D)$
\textcolor{black}{ and $\lim_{x\uparrow+\infty}\mu_2'(x)\ge 0$.} 
\\
We distinguish between two cases for the parameter $\nu$: 
\vspace{2mm}

\HYP{s1}: $\nu\in(1/2,1)$ and $\mu_1(0)>0$.\\
\HYP{s2}: $\nu=1/2$ and there exists $\bar{x}>0$ such that $2\mu_1(x)/\gamma^2\geq1$ for all $0 < x < \bar{x}$.

We now prove a rate of convergence as a corollary of Theorem~\ref{thm:Main}.

\begin{Proposition}[Locally smooth coefficients]\label{corollary:locallysmooth}
Assume that \HYP{s0} holds. Then,
$$\max_{i=0,\ldots,n}\left( \|\delta Y_{i}\|_2 + \|\delta X_{i}\|_1 + {\|\delta \breve{X}_{i}\|_{1+\epsilon}^{1+\epsilon}} \right)\le {C_{r,\epsilon}h^{r}}, \; \epsilon \ge 0,$$ 
with 
\begin{enumerate}
\item If \HYP{s1} holds, $r=1$.
\item If \HYP{s2} and $2\mu_1(0)/\gamma^2 =: \omega >3$ hold, $r\in(\frac16,1/2-1/\omega)$ if $3<\omega\leq 4$, $r=1/2$ if $4< \omega \leq 6$ and $r=1$ if $\omega>6$.
\end{enumerate}
{In both cases, we set $\zeta := h^{-\frac{2r}{q'-2}}$, with $q'=3+4\epsilon$ in the definition of $\breve{X}=\breve{Y}^2$, recall Remark~\ref{rem:dif schemes}.}
\end{Proposition}

\proof
In~\cite[Proposition~3.1]{de2011smoothness}, 
De Marco proves that under \HYP{s0}, there exists a unique strong solution to~\eqref{eqn:musigmasde}, 
which stays in $[0,\infty)$ almost surely.
In addition, he shows that \HYP{s1} and \HYP{s2} further imply that $\P(\tau_0= \infty) = 1$,
where $\tau_0$ is the first time the process $X$ reaches zero.
We recall that once we perform the Lamperti transformation, the diffusion function is a constant.

We divide the proof in several parts: in (i) we show that the drift function $f$ is one-sided Lipschitz continuous;
in (ii) we show that $f$ is locally Lipschitz continuous, and hence conclude that~\eqref{eq f loc lip} and~\eqref{eq f 1-sided lip} hold. 
\textcolor{black}{This is based on a direct  study of $f'$ the derivative of $f$.}

(i) From~\eqref{eqn:flocallysmooth}, it follows that, for all $x  \in D$,
\textcolor{black}{
\begin{align}\label{eq exp deriv}
f'(x) = \tilde{\mu}_1'(x) &- \frac{\nu}{a} \tilde{\mu}_1(x)x^{-\frac1{1-\nu}} 
- a\tilde{\mu}'_2(x)x^{\frac1{1-\nu}}
-(1-\nu) \tilde{\mu}_2(x) + \frac{\nu}{2(1-\nu)}x^{-2}\;,
\end{align}
where $a=[\gamma(1-\nu)]^{\frac1{1-\nu}}$ and for $g=\mu_1', \mu_1, \mu_2' \text{ or } \mu_2$, we set $\tilde g (x):= g \circ F^{-1}(x) = g(ax^{\frac1{1-\nu}})$, for all $x \in D$.\\
If $\nu \in (\frac12,1)$, under \HYP{s0} and \HYP{s1}, we have that $\lim_{y\rightarrow +\infty}f'(y) =-\infty$ (since $\mu'_2 \ge 0$) and $\lim_{y\rightarrow 0}f'(y) =-\infty$ as well (since $\mu_1(0)>0$ and $\nu>\frac12$).\\
If $\nu = \frac12$, under \HYP{s0} and \HYP{s2}, we deduce from the same arguments as previously that $\lim_{y\rightarrow +\infty}f'(y) =-\infty$. In this case, we obtain
$\lim_{y\rightarrow 0}f'(y) =-\infty$ because $\frac{\mu(0)}{\gamma^2} \ge \frac14$.
}
(ii) We now show that $f$ is locally Lipschitz continuous. 
\textcolor{black}{
From $\eqref{eq exp deriv}$ and the boundedness assumptions on $\mu_1$, $\mu_2$,$\mu'_1$ and $\mu'_2$, we obtain
\begin{align*}
|f'(x)| \le  C(1 + x^{\frac1{1-\nu}} + x^{-\frac1{1-\nu}} + x^{-2})\;,\; \text{for all } x \in D\;.
\end{align*}
Observing that for $\nu \in [\frac12,1)$, $x^{-2} \le 1 +  x^{-\frac1{1-\nu}}$, for all  $x \in D$,
we obtain that $f$ is locally Lipschitz continuous, with $\alpha=\beta=1/(1-\nu)$. 
}

Combining this with~(i) allows us to conclude that~\eqref{eq f loc lip} and~\eqref{eq f 1-sided lip} hold.

We now prove statements $1$ and $2$ in the corollary.

1) Assume \HYP{s1}. 
Since the locally Lipschitz exponents are $\alpha=\beta=1/(1-\nu)$, 
fix $k=k'=(1-\nu)/2$, so that \HYP{p} holds. 
By~\cite{de2011smoothness}, $\mathbb{E}(\sup_{t \in [0,T]} |X_t^{p}|)$ 
and $\mathbb{E}(\sup_{t \in [0,T]} |X_t|^{-p})$ are finite for all $p>0$;
therefore 
$\mathbb{E} (\sup_{t \in [0,T]} |Y_t|^{-q})$ is finite for all $q>0$~\cite[Lemma 3.1]{de2011smoothness}.
{We note that $f$ belongs to the class $\mathcal{C}^2(D)$ and \HYP{y2} holds, therefore $r=1$ from Proposition~\ref{first order rate}. 
The proof of the statement for $\| \delta \breve{X}_i\|_{1+\epsilon}$ follows from the same arguments as in the proof of Corollary \ref{cor:cir L1eps norm}.}

2){ Assume that \HYP{s2} holds and let $2\mu_1(0)/\gamma^2  =: \omega> 3$. Here, $\alpha=0$ an $\beta=0$.
Then, $\max_{t\in[0,T]}\mathbb{E}(|X_t|^{-p})$ is finite for all $p< \omega-1$~\cite[Lemma 3.1]{de2011smoothness}, and so is $\max_{t\in[0,T]}\mathbb{E} (|Y_t|^{-\ell})$ for all $\ell<2(\omega-1)$. 
Fix $q \in (4,2(\omega-1))$ and set $k=1/(q+2)$, so that \HYP{p}  and \HYP{y1} hold. 
From Theorem~\ref{thm:Main}, $r=1/2-\beta/(q+2) \in (\frac16, \frac12-\frac1\omega)$ holds.
}

{Further assume that $4<\omega \leq 6$.}
Note that the drift function $f$ belongs to the class~$\mathcal{C}^2(D)$.
Fix $q \in (8,2\omega)$ and  $k=1/4$, so that~\HYP{p} holds.  
By the assumptions on the parameters it follows that $\max_{t\in[0,T]}\mathbb{E}  (|Y_t|^{-6})=\max_{t\in[0,T]}\mathbb{E} (|X_t|^{-3})$ is finite, and therefore~\HYP{y2} holds.
{From Theorem~\ref{thm:Main}, $r=\min(1/2,(q+2)/8 - 1/2) >1/2$.} 
{Finally, in the case $\omega>6$, we can apply Proposition~\ref{first order rate}, to conclude that $r=1$. \\
The proof of the statement for $\| \delta \breve{X}_i\|_{1+\epsilon}$ follows from the same arguments as in the proof of Corollary \ref{cor:cir L1eps norm}.}
\eproof
\vspace{2mm}

In the CIR model, we obtain $r=1/2$ for $3 < \omega < 5$, 
using finite inverse moments of the process~$Y$ from~\cite{dereich2012euler}. 
For the general case in Proposition~\ref{corollary:locallysmooth}, we assumed that $4<\omega<6$ for $r=1/2$. 

In the next corollary, we impose additional assumptions in order to recover the same parameter constraints as for the Feller diffusion in the previous section.

\begin{Proposition}
Assume \HYP{s0} and \HYP{s2}. Moreover, let $a^*,b^*>0$ be such that
$\mu_1(x)\geq a^*$ and $\mu_2(x)\leq b^*$ for all $x \in D = (0,\infty)$.  
Then, 
$$\max_{i=0,\ldots,n} \left(\|\delta Y_{i}\|_2 + \|\delta X_{i}\|_1 + \| \delta \breve{X} \|_{1+\epsilon}^{1+\epsilon} \right) \le C_{r,\epsilon}h^{r},\; \epsilon \ge 0\,,$$
with $r=1/2$ if  $3 < \omega := 2 \mu_1(0)/\gamma^2 \leq 5$, and $r=1$ if $\omega>5$.\\ 
We set $\zeta := h^{-\frac{2r}{q'-2}}$, with $q'=3+4\epsilon$ in the definition of $\breve{X}=\breve{Y}^2$, recall Remark~\ref{rem:dif schemes}.
\end{Proposition}

\proof
From the assumptions on $\mu_1$ and $\mu_2$, there exists $a^*, b^*>0$ 
such that the inequality $\mu_1(x) - \mu_2(x) x \geq a^* - b^* x$ holds 
in the domain~$D$. 
We define $Z$ as the process with drift $a^* - b^* x$ (instead of $\mu_1(x)-\mu_2(x)x$), 
and diffusion $\sigma(x) \equiv \gamma x^{1/2}$. 
Therefore, by the Comparison Theorem (see~\cite[Section~5.2]{karatzas}) 
the inequality $X_t \geq Z_t$ holds for all $t\in[0,T]$ almost surely,
and hence $\mathbb{E} (|X_t|^{-p}) \leq \mathbb{E} (|Z_t|^{-p})$ is true for all $p>0$.
Now, $Z$ is clearly a Feller diffusion and, from the assumption on~$\omega$, it follows that 
$\max_{t\in[0,T]}\mathbb{E} (|Z_t|^{-3})$ is finite. 
The result then follows directly from the second part of Corollary~\ref{cor:circonvergencenew}.\\
The proof of the statement for $\| \delta \breve{X}_i\|_{1+\epsilon}$ follows from the same arguments as in the proof of Corollary \ref{cor:cir L1eps norm}.
\eproof

\subsection{$3/2$ model}

The $3/2$ process $X=(X_t)_{t\geq0}$~\cite{hestonlectures} is the solution to
\begin{align}\label{eqn:32process}
\ud X_t = c_1 X_t (c_2- X_t) \ud t + c_3 X_t^{3/2} \ud W_t , \quad X_0 = x_0 >0 , 
\end{align}
with $c_1, c_2, c_3 >0$.
Introduce the quantity $\omega:=2+2c_1/c_3^2$.
The Feller diffusion and the $3/2$ process are related as follows:
the map $F(y)\equiv y^{-1/2}$ yields the Lamperti transformed CIR process $Y:=F(X)$, 
as in~\eqref{eqn:lamperticir} and~\eqref{eqn:lamperticir2}, 
with parameters, $a:=(4c_1+3c_3^2)/8$, $b:= -c_1 c_2/2$ and $c:=-c_3/2$. 
Existence and uniqueness can be retrieved from the properties of the Feller diffusion,
and $\max_{t\in[0,T]} \mathbb{E} (|X_t|^p)$ is finite for all $p < \omega$.

\begin{Corollary}[$3/2$ model]\label{cor:32modellamperti}
Let $Y := X^{-1/2}$.
{Then, $\max_{i=0,\ldots,n} \|\delta Y_{i}\|_2  \le C h^r$, with 
$r\in(\frac16,\frac12-\frac1{w+1})$ if $\omega \in (2,3]$, $r=1/2$ if $3<\omega\leq5$ and $r=1$ if $\omega>5$.}
\end{Corollary}

\proof
In terms of the CIR coefficients, we have $\omega =2+2c_1/c_3^2 = 2\kappa\theta/\xi^2$.
{We directly apply Corollary~\ref{cor:circonvergencenew}} to get the desired results. 
\eproof
\vspace{2mm}

We now establish a convergence result for the $3/2$ process $X$, using the modification~$\tilde{X}$ (recall  Remark~\ref{rem:dif schemes}).

\begin{Proposition}\label{cor 32 mod lp}
{
Let $\omega>3$ and fix $\varepsilon\geq0$. 
If $3+2\varepsilon<\omega$, then 
$$\max_{i=0,\ldots,n} \|X_{t_i} - \tilde{X}_{t_i}\|_{1+\varepsilon} 
\leq C_{r,\varepsilon} h^{\frac{r}{2(1+\varepsilon)}}, $$with 
$r=1/2$ for $\omega\leq5$ and $r=1$ for $\omega>5$, where $\eta=h^{r/(2\omega)}$. 
}
\end{Proposition}

\proof
{
It follows that
\begin{equation}
\begin{array}{rl}
\nonumber  \|X_{t_i} - \tilde{X}_{t_i}\|_{1+\varepsilon}^{1+\varepsilon} & = \esp{|X_{t_i} - \tilde{X}_{t_i}|^{1+\varepsilon}} \\
 & = \esp{|\frac{1}{Y_{t_i}^2} - \frac{1}{\tilde{Y}_{t_i}^2}|^{1+\varepsilon}} \\
 & = \esp{\bigg|\frac{(Y_{t_i}-\tilde{Y}_{t_i})(Y_{t_i}+\tilde{Y}_{t_i})}{Y_{t_i}^2 \tilde{Y}_{t_i}^2} \bigg|^{1+\varepsilon}}\\
 & \leq \|Y_{t_i} - \tilde{Y}_{t_i}\|_{2} \sqrt{\esp{\frac{(Y_{t_i}+\tilde{Y}_{t_i})^{2+4\varepsilon}}{|Y_{t_i}|^{4+4\varepsilon} \tilde{Y}_{t_i}^{4+4\varepsilon} } }} \\
 & \leq C_\varepsilon \|Y - \tilde{Y}\|_{2} \sqrt{  \esp{ \frac{1}{|Y|^2 \tilde{Y}^{4+4\varepsilon}} + \frac{1}{|Y|^{4+4\varepsilon} \tilde{Y}^2} }        } 
\end{array}
\end{equation}
\textcolor{black}{
where we used Young's inequality to obtain the last inequality. We now compute
\begin{align*}
  \|X_{t_i} - \tilde{X}_{t_i}\|_{1+\varepsilon}^{1+\varepsilon}
  &
  \leq C_\varepsilon \|Y_{t_i} - \tilde{Y}_{t_i}\|_{2} \sqrt{\esp{ |\tilde{Y}_{t_i} \wedge Y_{t_i}|^{-(6+4\varepsilon)}  }}
\\
& \leq C_\varepsilon \|Y_{t_i} - \tilde{Y}_{t_i}\|_{2} \sqrt{\esp{ |Y_{t_i}|^{-(6+4\varepsilon)} + |\tilde{Y}_{t_i}|^{-(6+4\varepsilon)} }}. 
\end{align*}
}
Since $3+2\varepsilon<\omega$ it follows that $\esp{ |Y_{t_i}|^{-(6+4\varepsilon)}}$ is bounded by a constant. 
Furthermore, for $\eta =  h^{r/(2\omega)}$ ($q$ is such that $q<2\omega$), it follows that
$\esp{|\tilde{Y}_{t_i}|^{-(6+4\varepsilon)}} \leq \eta^{-(6+4\varepsilon)}$, 
therefore $\sqrt{ \esp{|\tilde{Y}_{t_i}|^{-(6+4\varepsilon)}} } \leq C_{\varepsilon, \omega} h^{-r/2}$, 
which together with $3+2\varepsilon<\omega$ and Corollary \ref{co conv ext scheme} (or Proposition \ref{first order rate}, if $r=1$), conclude the result.  
}
\eproof

\textcolor{black}{
\begin{Remark} \label{comp lit 3/2}
The last corollary proves $L^p$-bounds ($p>1$) for the $3/2$ model and improves the existing literature by yielding strong rates of convergence for $\omega>3$. 
More specifically, in the $L^2$-case, Neuenkirch and Szpruch~\cite[Proposition~3.2]{neuenkirch2012first} shows a rate~$1$ of convergence using a drift-implicit scheme when $\omega>12$, 
and Sabanis~\cite[Theorem~2]{sabanis2013super} gives a rate~$0.5$ for
$\omega \in (6,\infty)$. 
Corollary~\ref{cor 32 mod lp} improves these to an $L^2$-rate of convergence also for $\omega \in (5,6]$.
\end{Remark}
}

Alternatively, we could indeed use Proposition~\ref{finite moments} for a higher rate of convergence, 
however the parameter~$\omega$ required is larger:

\begin{Corollary}\label{full order 32 mod}
Let $\omega>\frac{9+4\varepsilon}2\vee5$ for some fixed $\varepsilon\geq0$. 
Then 
$$\max_{i=0,\ldots,n} \|X_{t_i} - \tilde{X}_{t_i}\|_{1+\varepsilon} 
\leq C_{\varepsilon,\omega} h^{1/(1+\varepsilon)}.$$
\end{Corollary}
\proof
{
From the computation in the proof of Proposition~\ref{cor 32 mod lp}, we have
\begin{align*}
 \|X_{t_i} - \tilde{X}_{t_i}\|_{1+\varepsilon}^{1+\varepsilon} & 
 \leq C_\varepsilon \|Y_{t_i} - \tilde{Y}_{t_i}\|_{2} \sqrt{\esp{ |Y_{t_i}|^{-(6+4\varepsilon)} 
	  + \esp{|\tilde{Y}_{t_i}|^{-(6+4\varepsilon)}} }}; 
\end{align*}
Using Proposition~\ref{finite moments}(ii), the term $\esp{|\tilde{Y}_{t_i}|^{-(6+4\varepsilon)}}$ 
is bounded by a constant depending on $\omega$ and $\varepsilon$, since $6+4\varepsilon < q -3 < 2\omega-3$.
Moreover, since $\omega>5$, we get that $\|Y_{t_i} - \tilde{Y}_{t_i}\|_{2} \le Ch$, from \eqref{eqn:cir_r} and the
same arguments as in the proof of Proposition~\ref{first order rate}.
}
\eproof

\subsection{Ait-Sahalia model}
In the Ait-Sahalia interest rate model~\cite{ait1996testing}, $X$ is the solution to
\begin{align}
\ud X_t = \left(\frac{a_{-1}}{X_t} - a_0 + a_1 X_t - a_2 X_t^\varrho\right) \ud t + \gamma X_t^\rho \ud W_t ,  \quad X_0 = x_0 > 0 , \label{eqn:aitsahsde}
\end{align}
where all constant parameters are non-negative, and $\rho, \varrho > 1$. 
From~\cite{szpruch2011numerical}, there exists a strong solution on~$(0,\infty)$, 
and the Lamperti transformation $Y := X^{1-\rho}$ satisfies 
\begin{align}
\ud Y_t = f(Y_t) \ud t + (1-\rho)\gamma \ud W_t ,  \quad Y_0 = x_0^{1-\rho} > 0 ,
\end{align}
with
$$ f(x) \equiv (1-\rho) \left( a_{-1} x^{\frac{-1-\rho}{1-\rho}} - a_0 x^{\frac{-\rho}{1-\rho}} + a_1 x - a_2 x^{\frac{-\rho+\varrho}{1-\rho}} - \frac{\rho \gamma^2}{2} x^{-1} \right) \; .$$

\begin{Corollary}
{If $\varrho+1>2\rho$, then $ \max\limits_{i=0,\ldots,n} \|\delta Y_{i}\|_2 
\le C h$.}
\end{Corollary}

\proof
Straightforward differentiation yields
$$
f'(x) = -a_{-1} (1+\rho) x^\frac{2}{\rho-1} + a_0 \rho x^\frac{1}{\rho-1} + a_1 (1-\rho) - a_2 (-\rho+\varrho) x^{-\frac{r-1}{\rho-1}} - \frac{\rho \gamma^2}{2} (\rho-1) x^{-2}. 
$$
We have $\lim_{x\downarrow 0}f'(x) = \lim_{x\uparrow \infty} f'(x) = -\infty$, 
hence $\sup_{0<x<\infty} f'(x)$ is finite by continuity and therefore  $f$ is one-sided Lipschitz continuous. 
In addition, 
$|f'(x)| \leq C(1 + x^\frac{2}{\rho-1} + x^{-\frac{\varrho-1}{\rho-1}} )$ for $x>0$,
so $f$ is locally Lipschitz continuous with $\alpha=2/(\rho-1)$ and $\beta=(\varrho-1)/(\rho-1)$. 
The diffusion is constant, hence Lipschitz continuous. 
Using the locally Lipschitz continuous properties of the drift, fix $k=1/(2\beta)$ and $k' = 1/(2\alpha)$. 
{We recall that if $\varrho+1>2\rho$, 
then $\max_{t\in[0,T]} \mathbb{E}(|X_t|^p)$ and $\max_{t\in[0,T]} \mathbb{E}(|X_t|^{-p})$ 
are finite for all $p\neq0$~\cite[Lemma~2.1]{szpruch2011numerical} so that
\HYP{y1} holds.}
Differentiation yields 
$$ f''(x) =  \frac{-2 a_{-1}(\rho+1)}{\rho-1} x^\frac{3-\rho}{\rho-1} + \frac{a_0 \rho}{\rho-1} x^\frac{2-\rho}{\rho-1} + a_2 \frac{(-\rho+\varrho)(\varrho-1)}{\rho-1} x^{-\frac{\varrho+\rho-2}{\rho-1}} + \rho \gamma^2 (\rho-1) x^{-3} \; . $$
Since $f$ belongs to  $\mathcal{C}^2(D)$ and~\eqref{eqn hy2 ass} is finite by~\cite[Lemma~2.3]{szpruch2011numerical}, then \HYP{y2} holds.

Fix $q>6\beta-2 $ and $q' > 6\alpha+2$. 
Then, by Proposition~\ref{first order rate}, the statement is proved.
\eproof
\vspace{2mm}

We now compute a strong rate of convergence for the Ait-Sahalia process $X$. We need to control the behaviour of the approximation near $0$ and
at $\infty$. In order to do that, we introduce modification $\check{X}_{t_i} := \check{Y}^{\frac1{1-\rho}}$ where $\check{Y}_{\ti{}} = p_{\bar{D}_\eta}\circ p_{\breve{D}_\zeta} (\hat{Y}_{t_i}) = p_{\check{D}_{\eta,\zeta}} (\hat{Y}_{t_i})$, 
for $\eta$ and $\zeta$ to be determined later on. 
\begin{Corollary} \label{coro conv aitsahalia}
If $\varrho+1>2\rho$, then for $\epsilon\ge 0$, 
$$\max\limits_{i=0,\ldots,n} \|X_{t_i} - \check{X}_{t_i}\|_{1+\epsilon} \leq Ch^{\frac1{1+\epsilon}}\,$$ 
with  $\eta:=h^{2/q}$, $\zeta=h^{-\frac{2}{q'-2}}$ and $q=3+4\rho(1+\epsilon)/(1-\rho)$, $q'=4\epsilon+1$. 
\end{Corollary}

\proof
A similar approach to Proposition~\ref{finite moments} yields
$$ \E[|\delta \check{X}_{t_i} |^{1+\epsilon} ]\leq C \left(\mathbb{E}\left[|Y_{t_i}|^{4\rho(1+\epsilon)/(1-\rho)} + |Y_{t_i}|^{4\epsilon}
 +
 |\check{Y}_{t_i}|^{4\rho(1+\epsilon)/(1-\rho)}
 + |\check{Y}_{t_i}|^{4 \epsilon}
  \right]\right)^{\frac12}(\E|\delta \check{Y}_{t_i} |^2)^{\frac12}, $$
  where $\delta \check{X}_{t_i}=X_{t_i} - \check{X}_{t_i}$ and $\delta \check{Y}_{t_i}=Y_{t_i} -\check{Y}_{t_i}$.
Since $\rho>1$ and $\varrho+1>2\rho$, $\mathbb{E}[|Y_{t_i}|^{4\rho(1+\epsilon)/(1-\rho)}+ |Y_{t_i}|^{4\epsilon}]$ is finite.
Observing that $\check{Y} \le \breve{Y} + \eta$, $\frac1{\check{Y}}\le \frac1{\tilde{Y}} + \frac1{\zeta}$ and using Proposition \ref{finite moments},
we get $\esp{|\check{Y}_{t_i}|^{4\rho(1+\epsilon)/(1-\rho)}
 + |\check{Y}_{t_i}|^{4 \epsilon} } \le C$. Also, we compute
\begin{align*}
|Y_{\ti{}}-\check{Y}_{\ti{}}| &\le |Y_{\ti{}} - p_{\bar{D}_\eta}(Y_{\ti{}})| + |p_{\bar{D}_\eta}(Y_{\ti{}})- p_{\bar{D}_\eta}\circ p_{\breve{D}_\zeta} (\hat{Y}_{t_i})|
\\
&\le |Y_{\ti{}} - p_{\bar{D}_\eta}(Y_{\ti{}})| + |Y_{\ti{}} -p_{\breve{D}_\zeta}(Y_{\ti{}})| + | Y_{\ti{}}- \hat{Y}_{t_i}|\,
\end{align*}
recalling that $p_{\bar{D}_\eta}$ and $p_{\breve{D}_\zeta}$ are $1$-Lipschitz. Using similar arguments as in the proof of Corollary \ref{co conv ext scheme}, we then obtain
$(\E|\delta \check{Y}_{t_i} |^2)^{\frac12} \le Ch$
and the result follows.
\eproof

\textcolor{black}{
\begin{Remark}\label{re ait sahalia}
In \cite{neuenkirch2012first}, the authors prove order one $L^2$-rate of convergence for the same range of parameter
but using an implicit scheme.
\end{Remark}
}

\section{Numerical results}\label{sec:numerics}
In this section, we numerically confirm the strong convergence rate of the modified Euler scheme
for the CIR model, the one-dimensional stochastic Ginzburg-Landau equation with multiplicative noise, 
and the Ait-Sahalia model. 
For a process $X$, denote by $\hat{X}_{T}^{(j)}$ the modified Euler-Maruyama approximation at time~$T$ and~$X_{T}^{(j)}$ the closed-form solution (or reference solution), using the same Brownian motion path (the~$j^{\text{th}}$ path). 
The empirical average absolute error $\mathcal{E}$ is defined by 
$$\mathcal{E} := \frac{1}{M}\sum_{j=1}^M |X_{T}^{(j)} - \hat{X}_{T}^{(j)}|,$$
over $M$ sample paths, which we will set to $M=10000$. 
An equidistant time grid is used, with step sizes $h:=T/2^N$, for different values of~$N$. 
The strong error rates are computed by plotting $\mathcal{E}$ against the number of discretisation steps on a log-log scale, 
and the strong rate of convergence $r$ is then retrieved using linear regression. 

\subsection{CIR model}
\label{sub se cir}
The Lamperti-transformed drift-implicit square-root Euler method (see~\cite{dereich2012euler,neuenkirch2012first}) has a unique strictly positive solution defined for $i=0,\ldots, n-1$ by
$$Y_{t_{i+1}} = \frac{Y_{t_i} + c \Delta W_{i+1}}{ 2(1-b h_{i+1})} + \sqrt{\frac{(Y_{t_i} + c \Delta W_{i+1})^2}{ 4(1-b h_{i+1})^2} + \frac{a h_{i+1}}{1- b h_{i+1}}}, \qquad Y_0 = \sqrt{x_0}>0,$$
with $a, b, c$ defined in~\eqref{eqn:lamperticir2}. 
The CIR/Feller diffusion is recovered by setting $X_{t_i} = Y_{t_i}^2$ for $i\le n$, and 
we compare the modified explicit Euler scheme with this implicit scheme used as a reference solution (with a large number of time steps).

We compute the strong rates of convergence for the CIR process, 
where the implicit scheme is used as a reference solution. 
Set $(\kappa,\theta,\xi,T,x_0)=(0.125\omega,1, 0.5,1,1)$, such that $2\kappa\theta/\xi^2=\omega$. 
The cases $\omega=(1,1.5,2,2.5,3,3.5,4)$ are considered. 
The reference solution is computed using $N=12$. 
\begin{figure}[h!]
 \centering
 \includegraphics[width=9cm]{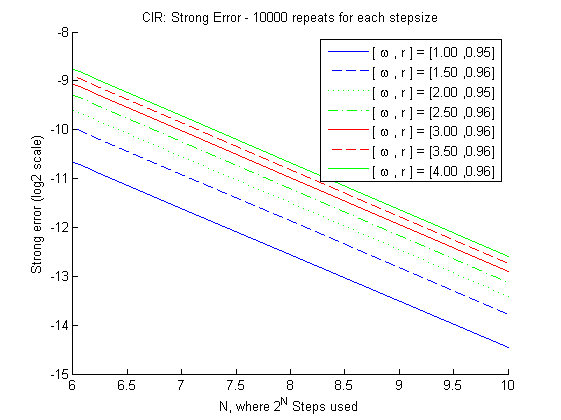}
 \caption{CIR model: $\mathcal{E}$ against number of steps ($\log_2$ scale).}
 \label{figure:cir_mod}
\end{figure}
Figure~\ref{figure:cir_mod} shows the rates of convergence $r$ achieved for the CIR process, where $k=1/4$ in the modified scheme, according to Corollary~\ref{cor:circonvergencenew}. 
In the corollary, we prove a strong rate of convergence of $1/2$ when $3<\omega\leq5$, and $r=1$ for $\omega>5$. 
The coefficient of determination $R^2$, {for the goodness of the fit of the straight line,} is above $0.998$ for all $\omega$. 
We observe that numerically order 1 is achieved by our scheme for $\omega>1$, which is better than the bound we proved.

\begin{Remark}\label{scaled proj}
The projection introduced in Definition~\ref{de scheme} can be modified to $\tilde{p}_n(x) :=  L n^{-k}\vee x \wedge U n^{k'}$,
with $L,U>0$ suitably chosen constant. 
This is beneficial if the process has extreme initial conditions or average state, and does not impact the convergence results.
\end{Remark}

\begin{figure}[h!]
 \centering
 \includegraphics[width=10cm]{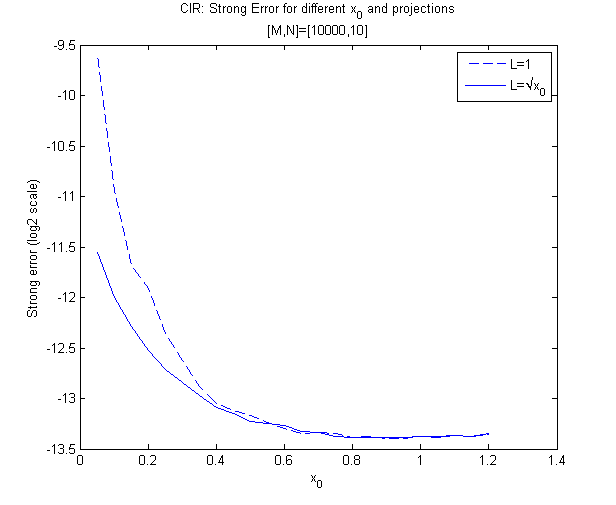}
 \caption{Absolute error ($\log_2$ scale) for $N=10$.}
 \label{figure:cir_proj_x0}
\end{figure}

For small $x_0$, it is intuitive to use the projection in Remark~\ref{scaled proj} to achieve faster convergence 
(albeit without affecting the asymptotic behaviour). 
Set $(\kappa,\theta,\xi,T)=(0.375,1, 0.5,1)$, such that $2\kappa\theta/\xi^2=3$. 
In Figure~\ref{figure:cir_proj_x0}, we let $x_0$ vary between $0.05$ and $1.2$ in increments of $0.05$. 
We compare the errors achieved for $k=1/4$, using the projections~$p_n(x)=n^{-k}\vee x$  and $\tilde{p}_n(x)=\sqrt{x_0}n^{-k}\vee x$.   
By using the projection $\tilde{p}_n$, smaller errors can be achieved for small $x_0$.  

\subsection{Ginzburg-Landau equation}
Consider the one-dimensional stochastic Ginzburg-Landau SDE~\cite[Chapter~4]{kloeden1992numerical}, where the process $X$ is the unique strong solution to 
$$ 
\ud X_t = \left[- X_t^3+ \left( \lambda+ \frac12 \sigma^2 \right) X_t \right] \ud t + \sigma X_t \ud W_t, 
\quad X_0 = x_0 > 0 ,
$$
for $\lambda,\sigma\geq0$, which admits the closed-form solution
\begin{equation}\label{eqn closed form gl}
 X_t = \frac{x_0 \exp(\lambda t + \sigma W_t)}{\sqrt{1+2x_0^2 \int_0^t \exp(2\lambda s + 2 \sigma W_s ) \ud s }}\;.
\end{equation}
This SDE is a special case of the Ait-Sahalia process with $(a_{-1},a_{0},a_1,a_2, \varrho,\rho)=(0,0,\lambda+\sigma^2/2,1,3,1)$.
For this choice of parameters, $\varrho+1>2\rho$, hence the moments and inverse moments of $X_t$ are finite for all $t\in[0,T]$, 
and the solution stays in $(0,\infty)$ almost surely. 
The drift function satisfies~\eqref{eq f loc lip}, with $(\alpha,\beta)=(2,0)$, e.g. set $k'=1/4$ in the modified scheme.
In addition, the drift is one-sided Lipschitz continuous and the diffusion is $K$-Lipschitz. 
As a result, theoretical convergence for this example can be obtained with rate 
$r=1$, recall also Remark \ref{re tamed}. 

\begin{figure}[h!]
 \centering
 \includegraphics[width=9cm]{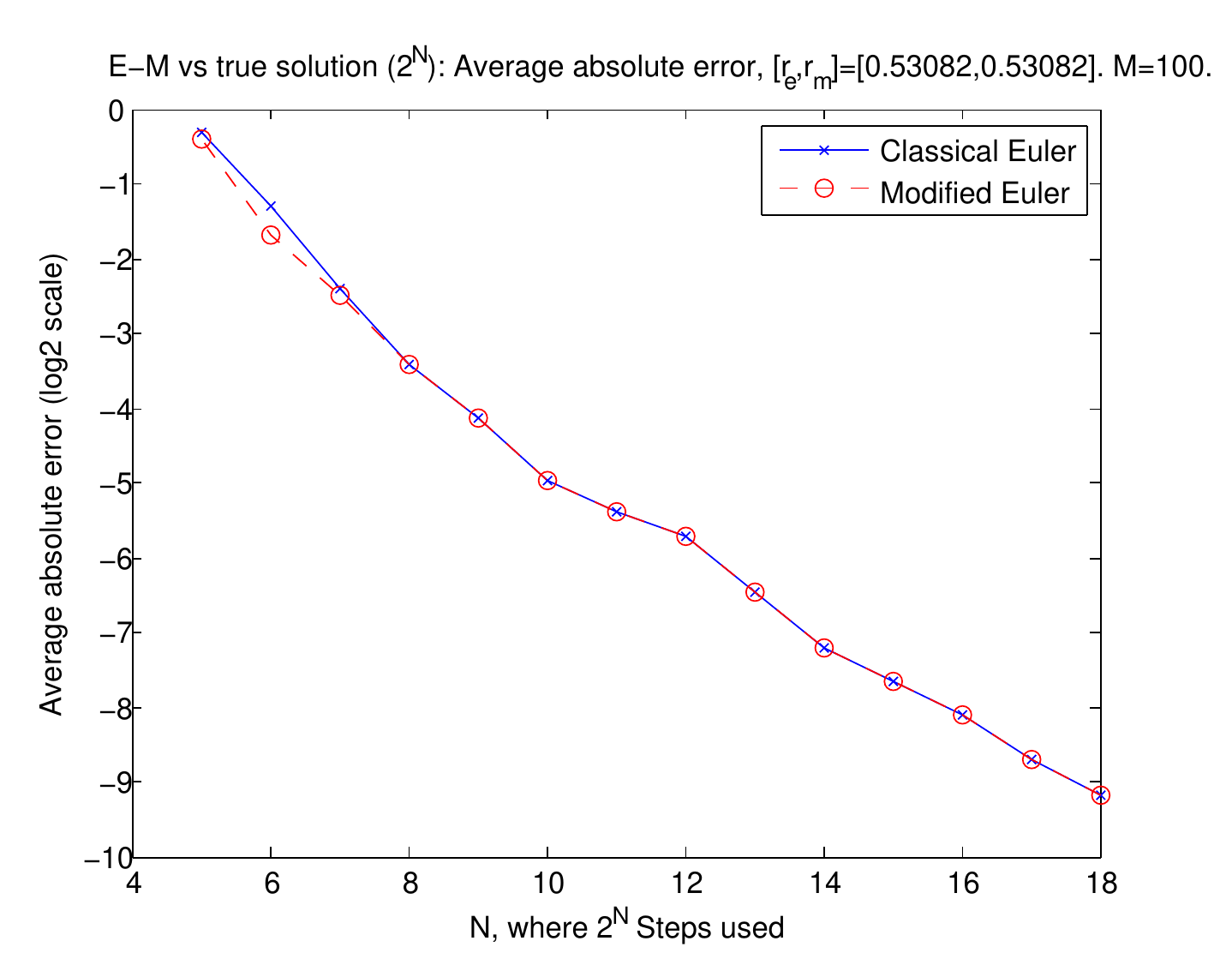}
 \caption{Ginzburg-Landau model: average absolute error $\mathcal{E}$ vs $N$ ($\log_2$ scale).}
 \label{figure:gl_numerics_emprojection_vs_sde}
\end{figure}

\paragraph{Ginzburg-Landau strong convergence:}
For this SDE, the closed-form solution is used in the definition of $\mathcal{E}$ to compute the strong rate of convergence $r$. 
Figure~\ref{figure:gl_numerics_emprojection_vs_sde} shows the average absolute error $\mathcal{E}$ using the modified scheme, for parameters $(\sigma, \lambda, T, x_0) =(1,1/2,1,1)$. 
The empirical rate achieved of $0.53$ (same as the standard Euler scheme) which is lower than the predicted rate of $1$. 
This can be explained since we are approximating the integral in~\eqref{eqn closed form gl} as a summation. 

\paragraph{Ginzburg-Landau Euler-Maruyama divergence:}
We consider an example of the Ginzburg-Landau SDE for which the standard Euler-Maruyama scheme diverges, and compare the results with the modified explicit scheme. 
Fix parameters~$(\sigma, \lambda, T, x_0) =(7,0,3,1)$ as in~\cite{hutzenthaler2011strong}, for which the  
authors prove moment explosion for the classical Euler-Maruyama scheme, see~\cite[Table~1]{hutzenthaler2011strong}. 
\begin{figure}[h!]
 \centering
 \includegraphics[width=10cm]{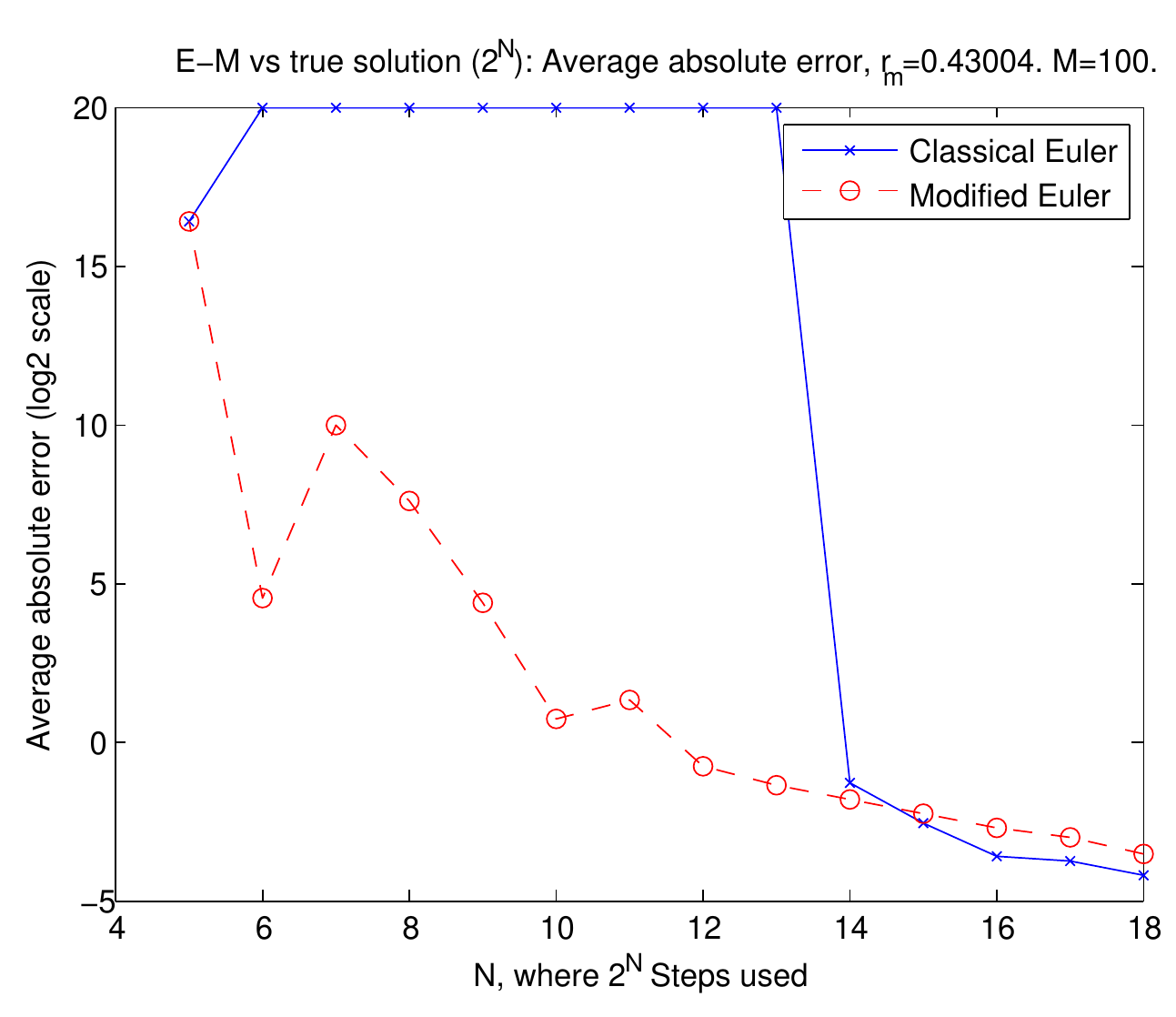}
 \caption{Average absolute error $\mathcal{E}$ vs number of steps ($\log_2$ scale).}
 \label{figure:gl_explosion_em_vs_sde}
\end{figure}
Figure~\ref{figure:gl_explosion_em_vs_sde} shows the error $\mathcal{E}$ for the classical and the modified schemes, for different $N$. 
For the modified scheme, set $k'=1/4$. 
The modified Euler scheme converges with a rate $r_m =0.43$. 
For a range of step sizes, the classical Euler scheme explodes, as proven in~\cite{hutzenthaler2011strong} (N.B. very large and $NaN$ values are set to $2^{20}$ in the figure, to illustrate the explosions for the classical scheme).
The modified scheme appears to be more robust.

\subsection{Ait-Sahalia model}
\begin{figure}[h!]
 \centering
 \includegraphics[width=10cm]{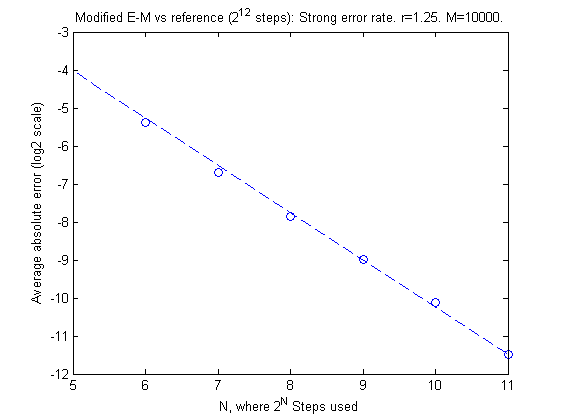}
 \caption{Ait-Sahalia model: average absolute error vs $N$ ($\log_2$ scale).}
 \label{figure:aitsahalia_numerics_emprojection_vs_reference}
\end{figure}

The strong rate of convergence for the Ait-Sahalia model is computed using a reference solution with a large number of steps. 
Consider the parameters $(a_{-1}, a_0, a_1, a_2,\gamma, x_0 ) = (1, 1, 1, 1,1,1)$, and $(\varrho, \rho, T) = (2, 3/2, 1)$.  
From these parameters, note that $\alpha = 4$ and $\beta = 2$. 
Fix $k$ and $k'$, such that $2\beta k=1$ and $2\alpha k' = 1$, so that~\HYP{y1} holds. 
Figure~\ref{figure:aitsahalia_numerics_emprojection_vs_reference} shows $\mathcal{E}$ against the number of steps (log-log plot), where $2^{12}$ steps are used for the reference solution. 
The Ait-Sahalia  empirical rate of convergence $r=1.25$ could be justified by the fact that we used a reference solution instead of the true solution.

\subsection{MLMC}
We combine the modified Euler scheme and the multilevel Monte Carlo approach introduced by Giles~\cite{giles2008multilevel,giles2012multilevel}. 
The original paper focused on approximating the expected value of Lipschitz continuous payoffs. 
The MLMC method has also been justified for digitals, lookback and barrier options~\cite{giles2009analysing}.
Multischeme MLMC techniques use different discretisation schemes in order to further improve the computational efficiency~\cite{abe2011pricing}.
The use of MLMC techniques has also been applied to compute Greeks~\cite{burgos2012computing}.

We target a root mean squared error~(RMSE) of $\mathcal{O}(\varepsilon)$ for the option price. 
Using an Euler-Maruyama scheme, the MSE of an option price is $C_1/N + C_2 h^2$, where $N$ is the number of Monte Carlo paths, and $h$ is the step size of the discretisation.
By choosing $N:=\mathcal{O}(\varepsilon^{-2})$, and $h:=\mathcal{O}(\varepsilon)$, the total cost is $\mathcal{O}(\varepsilon^{-3})$.

The idea behind MLMC is to use different time steps, at different levels of the simulation. 
We increase the number of time steps at each level by a factor~$M$, where level $l$ uses $M^l$ steps of size $h_l := T/M^l$. 
We define $P_l$ to be the numerical approximation of the payoff at level $l$, for $l=0,\ldots, L$, where $L$ is the maximum number of levels. 
By linearity of the expectation operator we note that 
\begin{align}
\mathbb{E} \left[ P_L \right] = \mathbb{E} \left[ P_0 \right] + \sum_{l=1}^L \mathbb{E} \left[ P_l - P_{l-1} \right] \; ,
\end{align}
where the difference in the payoff approximation on levels $l$ and $l-1$ is estimated using the same Brownian path, for both levels. 
The variance of the payoff difference, $V_l := \mathbb{V}(P_l - P_{l-1})$, decreases quickly with increasing levels, and it has been shown that for European options with Lipschitz continuous payoffs, $V_l$ converges to zero twice as fast as the strong convergence rate of the scheme. 
At each level $l$, we simulate $N_l$ paths and estimate $\mathbb{E} \left[P_l - P_{l-1} \right]$. 
The multilevel estimator has variance $1/N_l\sum_{l=0}^L V_l$, and $N_l := C \sqrt{V_l h_l}$ minimises the computational cost~\cite{giles2008multilevel}, to achieve a RMSE of $\mathcal{O}(\varepsilon)$.
The strong convergence rate is required for the MLMC techniques, and the complexity theorem provides a general result for the computational cost of the MLMC method~\cite{giles2008multilevel}. 
MLMC methods have been shown to improve the computational efficiency using an Euler-Maruyama discretisation to $\mathcal{O}\left(\varepsilon^{-2} (\log\varepsilon)^2\right)$, and $\mathcal{O}(\varepsilon^{-2})$ for a Milstein scheme~\cite{giles2008multilevel,giles2008improved}.

\paragraph{CIR model ZCB:}

We consider the Cox-Ingersoll-Ross model~\eqref{eq:FellerSDE} for the process $(v_t)_{t\geq0}$; 
the price of a zero-coupon bond (ZCB) with maturity $T$, at time $t$, reads
$$ B(t,T) = \mathbb{E} \left[ \exp\left(-\int_{t}^T v_s \ud s \right) \bigg| \mathcal{F}_t \right],$$
which admits a closed-form solution~\cite{cir1985,brigo2007interest}. 
This solution at time zero is $B(0,T) = A\exp(-C v_0)$, where $\Lambda:=\sqrt{\kappa^2 + 2\xi^2}$ and 
$$A:= \left(   \frac{2\Lambda\exp\left[(\kappa+\Lambda)T/2\right]}{2\Lambda + (\kappa+\Lambda)(\exp{T\Lambda}-1)}     \right)^{2\kappa\theta/\xi^2}, \qquad C:= \frac{2(\exp(T\Lambda)-1)}{2\Lambda + (\kappa+\Lambda)(\exp(T\Lambda)-1)}.$$ 

\begin{figure}[h!]
 \centering
 \includegraphics[scale=0.6]{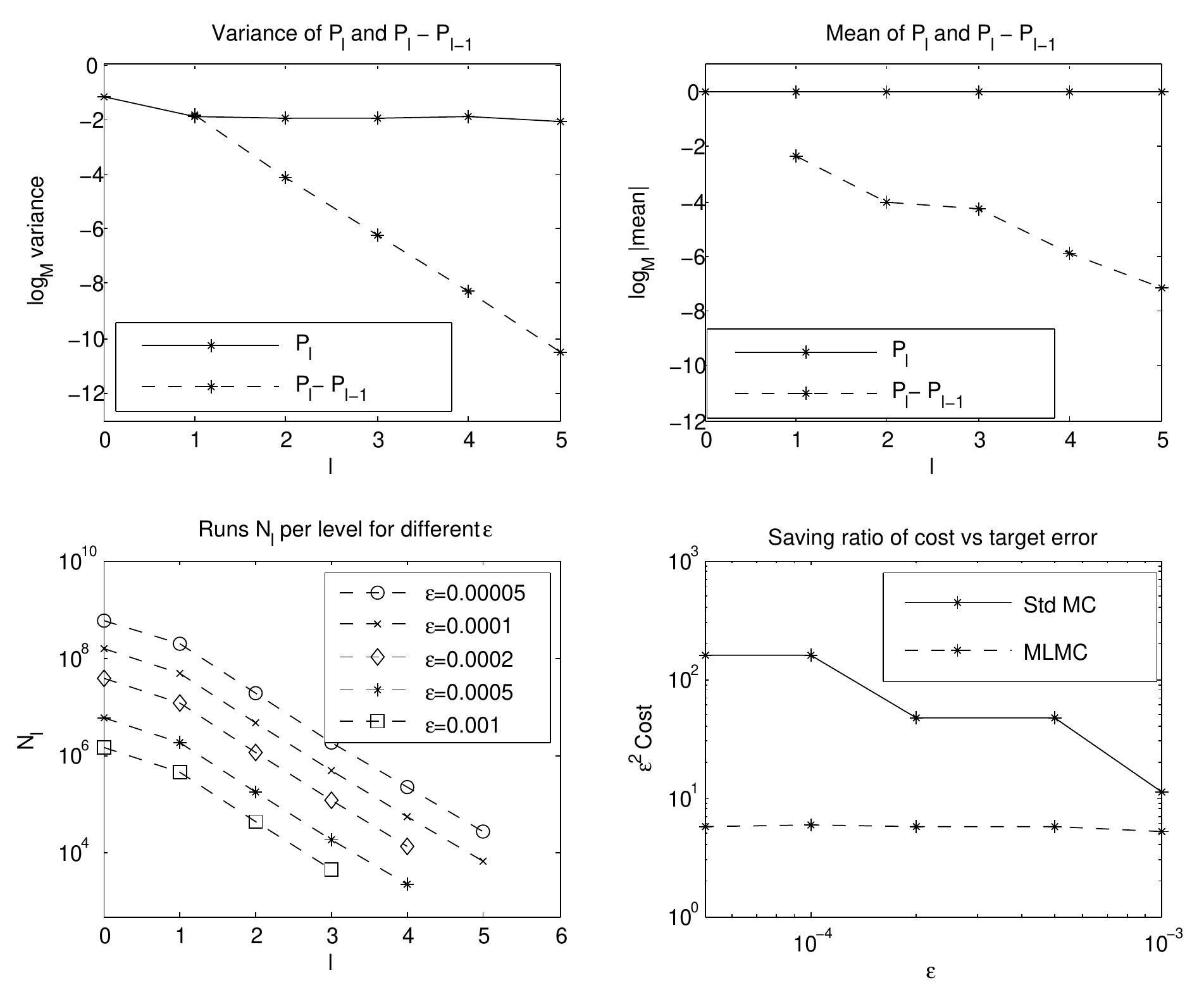}
 \caption{CIR model, and ZCB pricing using MLMC.}
 \label{figure:20140401_zcb_cir.eps}
\end{figure}
We consider a CIR model with parameters
$(\kappa, \theta, \xi, v_0, T) = (2,1,0.5,1,1)$, $(N,M,L) = (2000000,4,5)$, and RMSE thresholds $(0.001, 0.0005, 0.0002, 0.0001, 0.00005)$. 

In Figure~\ref{figure:20140401_zcb_cir.eps}, we compute the standard Monte Carlo, and MLMC approximations for the ZCB. 
The first plot demonstrates the average variance for the approximations $P_l$ and the differences $P_l - P_{l-1}$. 
Observe that the variance of the differences decreased roughly twice as fast as the rate of weak convergence of an Euler scheme. 
Also, the variance of $P_l$ is asymptotically a constant. 
The second plot shows the mean of $P_l$ and the mean of $P_l - P_{l-1}$. 
The third plot shows how decreasing the target $\varepsilon$ requires more steps $N_l$ and increases the number of levels from 3 to 5. 
The fourth plot shows the ratio of savings between the standard Monte Carlo approach for approximating the bond price (Std MC), and the MLMC counterpart. 
The ratio of savings is a factor of 27 for $\varepsilon=0.00005$ between the standard Monte Carlo and the MLMC approach. 
We adapt code freely available from~\cite{giles2008multilevel}.

\paragraph{CIR model spread option:}
We consider the CIR model for processes $(X^1_t)_{t\geq0}$ and $(X^2_t)_{t\geq0}$, the solutions of the following stochastic differential equations:

\begin{equation}\label{eq:2D-FellerSDE}
\begin{array}{rll}
\ud X_t^1 &= \kappa_1(\theta_1-X_t^1) \ud t + \xi_1 \sqrt{X_t^1} \ud W_t, 	&\qquad X_0^1 = x_0^1 > 0 , \\
\ud X_t^2 &= \kappa_2(\theta_2-X_t^2) \ud t + \xi_2 \sqrt{X_t^2} \ud Z_t,   	&\qquad X_0^2 = x_0^2 > 0 , \\
\ud W_t \ud Z_t &= \rho \ud t ,
\end{array} 
\end{equation}
where $W$, $Z$ are correlated Brownian motions with $-1\leq \rho\leq 1$ and $\kappa_1, \kappa_2, \theta_1, \theta_2, \xi_1, \xi_2$ are strictly positive constant parameters.  

The payoff of a European call spread option at the terminal time $T$ for a given strike~$K$ is defined as $\max(X_T^1-X_T^2-K,0)$. 
Spread options are used for hedging and speculation purposes and are widely traded in the commodity markets. 
Their price is particularly sensitive to the correlation parameter; for increasing $\rho$, the price of the spread option decreases.

\begin{Example}\label{eg:zero_rho_spread_option}
Suppose~\eqref{eq:2D-FellerSDE} with parameters $(\kappa_1, \theta_1, \xi_1)  = (1,0.06,0.04)$, $(\kappa_2, \theta_2, \xi_2)  = (0.8,0.05,0.016)$, $(x_0^1, x_0^2, \rho, T, K) = (0.05,0.06,0,1,0.001)$. 
We compute the option price using $N=10,000,000$ Monte Carlo paths using the implementation from~\cite[p.124]{glasserman2003monte}. 
The spread option price and its $95\%$ confidence interval is $0.00310063 \pm 0.00000267$. 
\end{Example}

\begin{figure}[h!]
 \centering
 \includegraphics[scale=0.6]{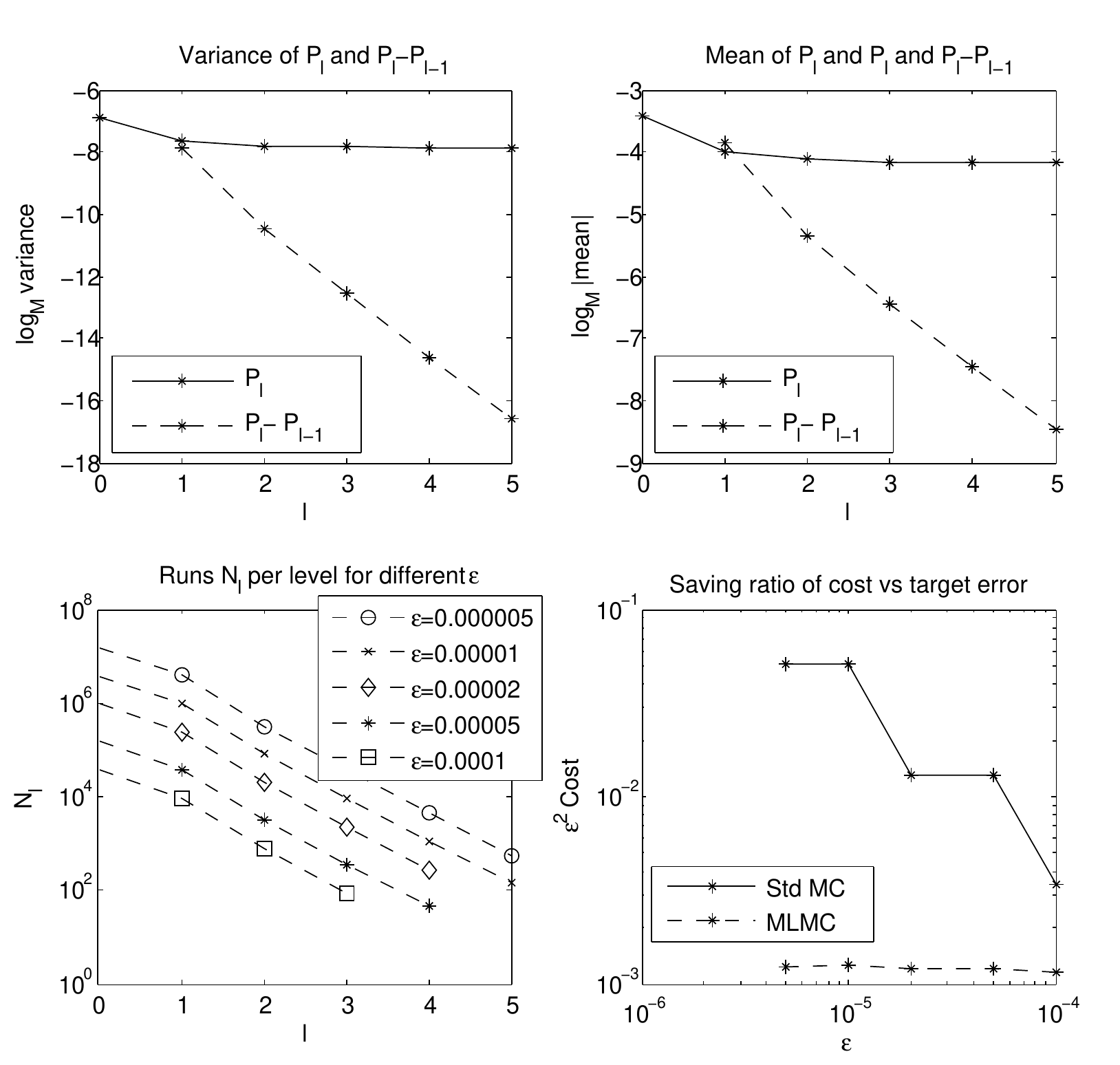}
 \caption{CIR spread option, $\rho=0$. MLMC for Example~\ref{eg:zero_rho_spread_option}.}
 \label{figure:20150414_spread_option_nocorr.pdf}
\end{figure}

We set $M=4$ throughout, where $M$ is the multiple of the step-sizes for the MLMC. 
In Table~\ref{tab:nocorr spread option}, the RMSE and the target $\varepsilon$ is shown. 
The savings column shows the speedup multiple of the MLMC computational cost compared to the standard Monte Carlo routine. 

\begin{table}[htbp]
  \centering
    \begin{tabular}{|c|c|c||c|}
\hline
    RMSE  & Target $\varepsilon$ & Ratio & Savings \\
\hline
    0.000046 & 0.0001 & 0.456   & 2.97 \\
    0.000032 & 0.00005 & 0.637 & 10.61 \\
    0.000018 & 0.00002 & 0.921 & 10.67 \\
    0.000007 & 0.00001 & 0.671   & 40.90 \\
    0.000004 & 0.000005 & 0.855 & 40.97 \\
\hline
    \end{tabular}%
  \caption{CIR spread option: RMSE and computational savings for Example~\ref{eg:zero_rho_spread_option}. }
  \label{tab:nocorr spread option}%
\end{table}%

\begin{Example}\label{eg:non-zero_rho_spread_option}
Suppose~\eqref{eq:2D-FellerSDE} with parameters $(\kappa_1, \theta_1, \xi_1)  = (1,0.06,0.04)$, $(\kappa_2, \theta_2, \xi_2)  = (0.8,0.05,0.016)$, $(x_0^1, x_0^2, \rho, T, K) = (0.05,0.06,-0.7,1,0.001)$. 
For this example, we compute the reference option price using the drift-implicit Euler scheme, using $N=10,000,000$ paths with $2^{12}$ time steps. 
The price and its $95\%$ confidence interval is $0.003711\pm0.0000032$. 
\end{Example}

For Example~\ref{eg:non-zero_rho_spread_option}, the RMSE, ratio to target $\varepsilon$ and savings factor over standard Monte Carlo are shown in Table~\ref{tab:corr spread option}.

\begin{figure}[h!]
 \centering
 \includegraphics[scale=0.6]{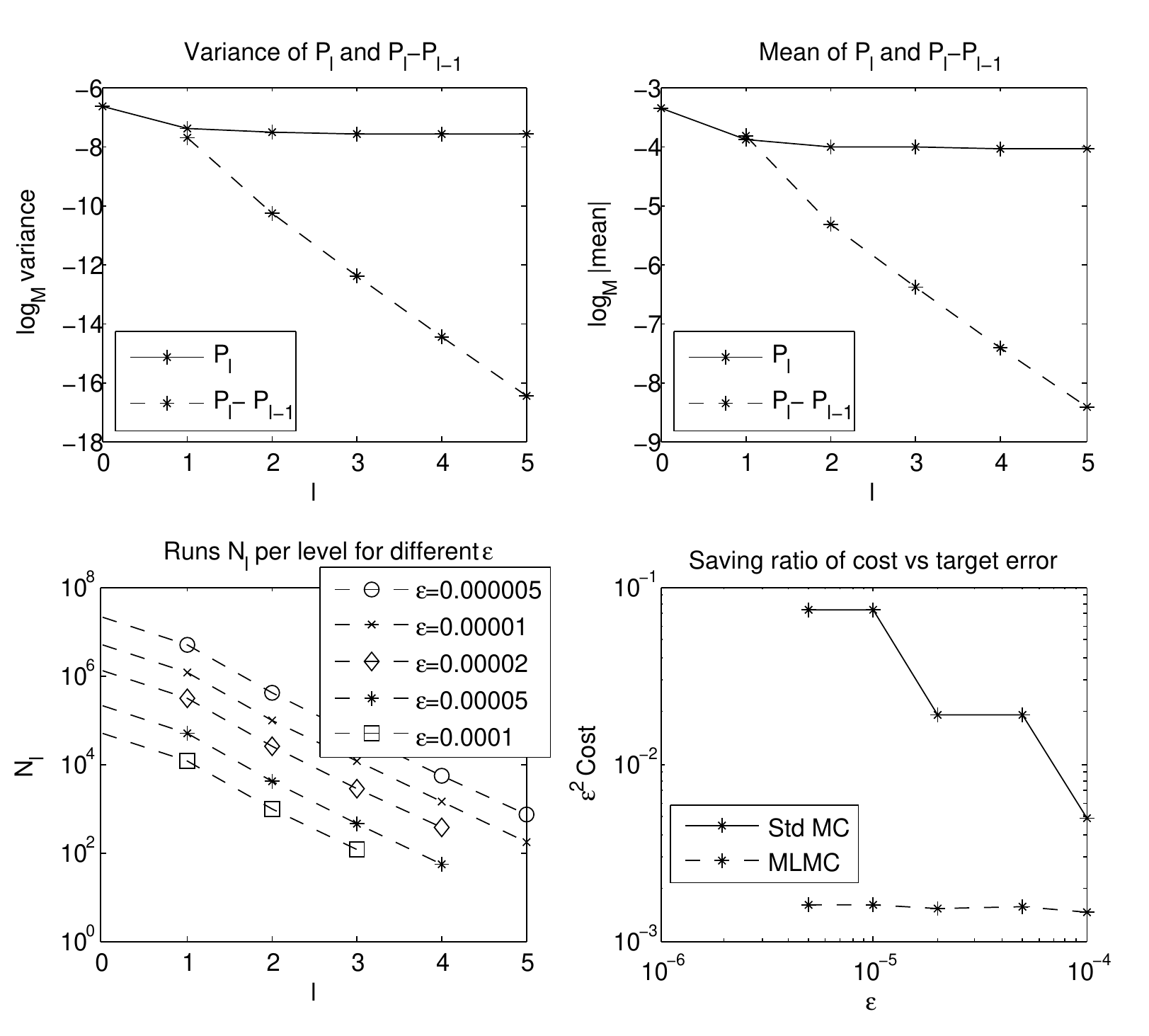}
 \caption{CIR spread option, $\rho=-0.7$. MLMC for Example~\ref{eg:non-zero_rho_spread_option}.}
 \label{figure:20150414_Spread_Neg_Corr_CIR.pdf}
\end{figure}

\begin{table}[htbp]
  \centering
    \begin{tabular}{|c|c|c||c|}
\hline
    RMSE  & Target $\varepsilon$ & Ratio & Savings \\
\hline
    0.000075 & 0.0001 & 0.751 & 3.39 \\
    0.000037 & 0.00005 & 0.745 & 12.15 \\
    0.000019 & 0.00002 & 0.953 & 12.2 \\
    0.000007 & 0.00001 & 0.707 & 47.2 \\
    0.000004 & 0.000005 & 0.811 & 47.19 \\
\hline
    \end{tabular}%
  \caption{CIR spread option: RMSE and computational savings for Example~\ref{eg:non-zero_rho_spread_option}. }
  \label{tab:corr spread option}%
\end{table}%

\begin{Remark}
For the above MLMC examples above, recall the constant~$L$ from Remark~\ref{scaled proj}, which for the CIR examples above is chosen at $0.01$. The modified Euler scheme parameter is set to $k=1/4$ as seen before. 
\end{Remark}

\renewcommand{\abstractname}{Acknowledgements}
\begin{abstract}
The authors would like to thank the anonymous referees for their suggestions, 
as well as Stefano de Marco and Lukasz Szpruch for valuable comments. 
Jean-Fran\c{c}ois Chassagneux acknowledges financial support from the ANR grant \textsc{Liquirisk}  ANR-11-JS01-0007. Antoine Jacquier acknowledges financial support from the EPSRC First Grant EP/M008436/1. 
\end{abstract}

\section*{References}
\bibliographystyle{amsplain}
\bibliography{ivom87}

\appendix
\section{Proof of Lemma~\ref{le fn lip}}\label{app:proof lemma}

{
The fact that $f_n$ is $L(n)$-Lipschitz continuous is straightforward. We prove 
the one-sided Lipschitz property in two steps below.\\
Step $1$. Let $r>l>0$ such that $D_n \subset (l,r)$. Assume that $f$ is $C^1(l,r)$. From {\eqref{eq f 1-sided lip}}, we have, for $z,z' \in D_n$, $z>z'$,
\begin{align*}
 \frac{f(z)-f(z')}{z-z'} \le K,
\end{align*}
and letting $z' \rightarrow z$, we retrieve that
$f'(z) \le K$.
This shows that $f= g + \ell$, where~$g$ is a non-increasing function and $\ell$ is $K$-Lipschitz continuous,
setting e.g. $g(x)\equiv\int_{\frac{l+r}2}^xf'(u)\1_{\set{f'(u)\le 0}} \ud u$ and $\ell(x)\equiv\int_{\frac{l+r}2}^xf'(u)\1_{\set{f'(u)> 0}} \ud u$.
Since $p_n$ is non-decreasing and $1$-Lipschitz on $\R$, we have $f_n = g \circ p_n + \ell \circ p_n$, with $g \circ p_n$ non-increasing and
$\ell \circ p_n$ $K$-Lipschitz continuous on $\R$. This shows that $f_n$ satisfies \eqref{eq f 1-sided lip} as well on $\R$.
\\
Step $2$. We now deal with the general case using a smoothing argument.
Let $l,r \in D$, $r>l$, such that for all $D_n \subset (l,r)$. 
We consider a sequence $(\varphi_m)_{m\ge 1}$ of mollifiers whose supports are included in $[-\frac{l}2,\frac{l}2]$ and define
$f^m \equiv \varphi_m \star f \equiv \int_{[-\frac{l}2,\frac{l}2]}\varphi_m(u)f(x-u)\ud u$ as the convolution of $\varphi_m$ and~$f$. 
We observe that, for all $x,y \in (l,r)$,
\begin{align*}
(x-y)(f^m(x) - f^m(y)) &= \int_{[-\frac{l}2,\frac{l}2]}\varphi_m(u)\{(x-y)(f(x-u)-f(y-u))\}\ud u
\\
&\le K|x-y|^2\int_{[-\frac{l}2,\frac{l}2]} \varphi_m(u) \ud u\;\le K|x-y|^2\;,
\end{align*}
where we used~\eqref{eq f 1-sided lip} and the fact that $\int_D\varphi_m(u) \ud u = 1$.
Since $f^m$ is smooth, we can apply Step $1$ to obtain, for all $(x,y) \in \R^2$,
\begin{align*}
 (x-y)\left(f^m(p_n(x)) - f^m(p_n(y))\right) \le K|x-y|^2\;.
\end{align*}
Letting $m$ go to infinity, we then obtain
\begin{align*}
 (x-y)\left(f(p_n(x)) - f(p_n(y))\right) \le K|x-y|^2\;,
 \end{align*}
for all $x,y \in \R$, which concludes the proof.
}

\end{document}